\documentclass[epjH]{svjour}
\usepackage{mathtext}
\usepackage[T1,T2A]{fontenc}
\usepackage[utf8]{inputenc}
\usepackage[ukrainian,english]{babel}
\usepackage[dvips]{graphicx}
\usepackage{wrapfig} 

\begin{document}
\title{Samuil Kaplan and the development of \\ astrophysical research at the Lviv University\\
(dedicated to the 100th anniversary of his birth)}
\author{B. Novosyadlyj$^{1}$\thanks{E-mail:bohdan.novosyadlyj@lnu.edu.ua}, B. Hnatyk$^2$, Yu. Kulinich$^1$, B. Melekh$^1$, O. Petruk$^3$, R. Plyatsko$^3$, M. Tsizh$^1$, \\ M. Vavrukh$^1$, N. Virun$^1$}
\institute{$^1$Ivan Franko National University of Lviv, 8, Kyryla i Methodia Str., Lviv, Ukraine, 79005; \\
$^2$Taras Shevchenko National University of Kyiv, 64/13, Volodymyrska Str., Kyiv, Ukraine, 01601; \\
$^3$Pidstryhach Institute for Applied Problems of Mechanics and Mathematics, NASU, 3-b, Naukova Str., Lviv, Ukraine, 79060}
\abstract{Samuil Kaplan (1921-1978) was a productive and famous astrophysicist. He was affiliated with a number of scientific centers in different cities of former Soviet Union. The earliest 13 years of his career, namely in the 1948-1961 years, he worked in Lviv University in Ukraine (then it was called the Ukrainian Soviet Socialist Republic). In the present paper, the Lviv period of his life and scientific activity is described on the basis of archival materials and his published studies. Kaplan arrived in Lviv in June 1948, at the same month when he obtained the degree of Candidate of science. He was a head of the astrophysics sector at the Astronomical Observatory of the University, was a professor of department for theoretical physics as well as the founder and head of a station for optical observations of artificial satellites of Earth. He was active in the organization of the astronomical observational site outside of the city. During the years in Lviv, Kaplan wrote more than 80 articles and 3 monographs in 9 areas. The focus of his interests at that time was on stability of circular orbits in the Schwarzschild field, on white dwarf theory, on space gas dynamics, and cosmic plasma physics, and turbulence, on acceleration of cosmic rays, on physics of interstellar medium, on physics and evolution of stars, on cosmology and gravitation, and on optical observations of Earth artificial satellites. Some of his results are fundamental for development of theory in these fields as well as of observational techniques. The complete bibliography of his works published during the Lviv period is presented. Respective scientific achievements of Samuil Kaplan are reviewed in the light of the current state of research in these areas.}
\maketitle
\section*{Introduction}
Samuil Aronovich Kaplan (10.10.1921 – 11.06.1978), a prominent astrophysicist of the twentieth century, began his active scientific career in Lviv at the age of 27 after graduating from the Leningrad (now St. Petersburg) State University in 1948. It was in Lviv that he wrote his first scientific papers and initiated studies in the field of theoretical astrophysics. On the occasion of the 100th anniversary of Samuil Kaplan’s birth, the Astronomical Observatory and the Department of Astrophysics of the Lviv University organized a one-day online seminar\footnote{The recording of the seminar is available at https://youtu.be/YewcivEhdss} on December 6, 2021, which laid basis for the publication of this paper. Among the speakers of the seminar, there were Samuil Kaplan’s ``scientific grandchildren and great-grandchildren'', scientists who are the contributors in his research fields from Lviv and other research centers both in Ukraine and abroad. In the present paper, we do not present a complete biography of the scientist as it is available on the Internet\footnote{https://en.wikipedia.org/wiki/Samuil\_Kaplan}; we decided to explore the Lviv period of his life, which received little coverage in the available sources. At the end of the paper, we include the bibliography of S.~Kaplan's works of the Lviv period, which impresses not only with the number of works (95 in 13 years) but also with the breadth of topics. Having analyzed his works, we can confidently claim that most of them are fundamental in the development of the theory of degenerated stars, stability of orbits in the gravitational field of the black holes, space gas dynamics, interstellar medium, and space plasma. S.~Kaplan’s research works have become classics of theoretical astrophysics.
 
\section{Pages from biography (based on the personal file and colleagues' memoirs)}
\begin{figure}[!t]
\includegraphics[height=0.26\textheight]{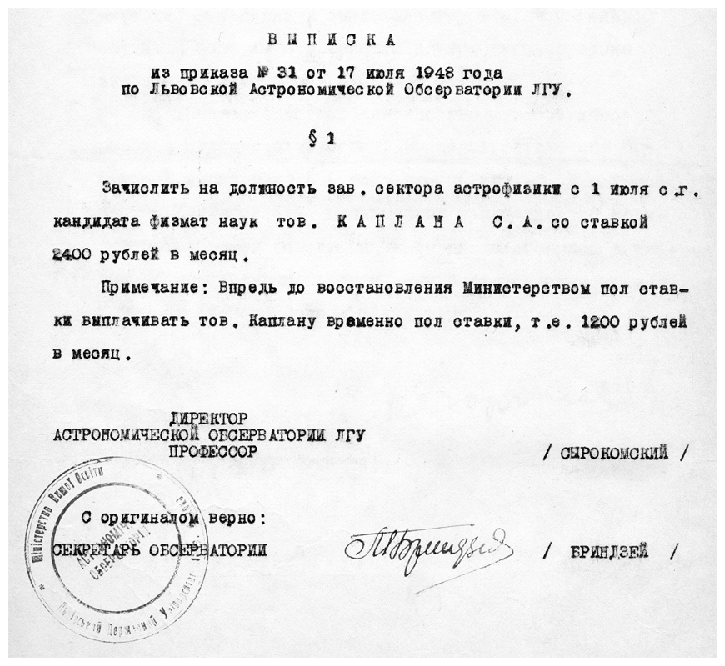}
\includegraphics[height=0.26\textheight]{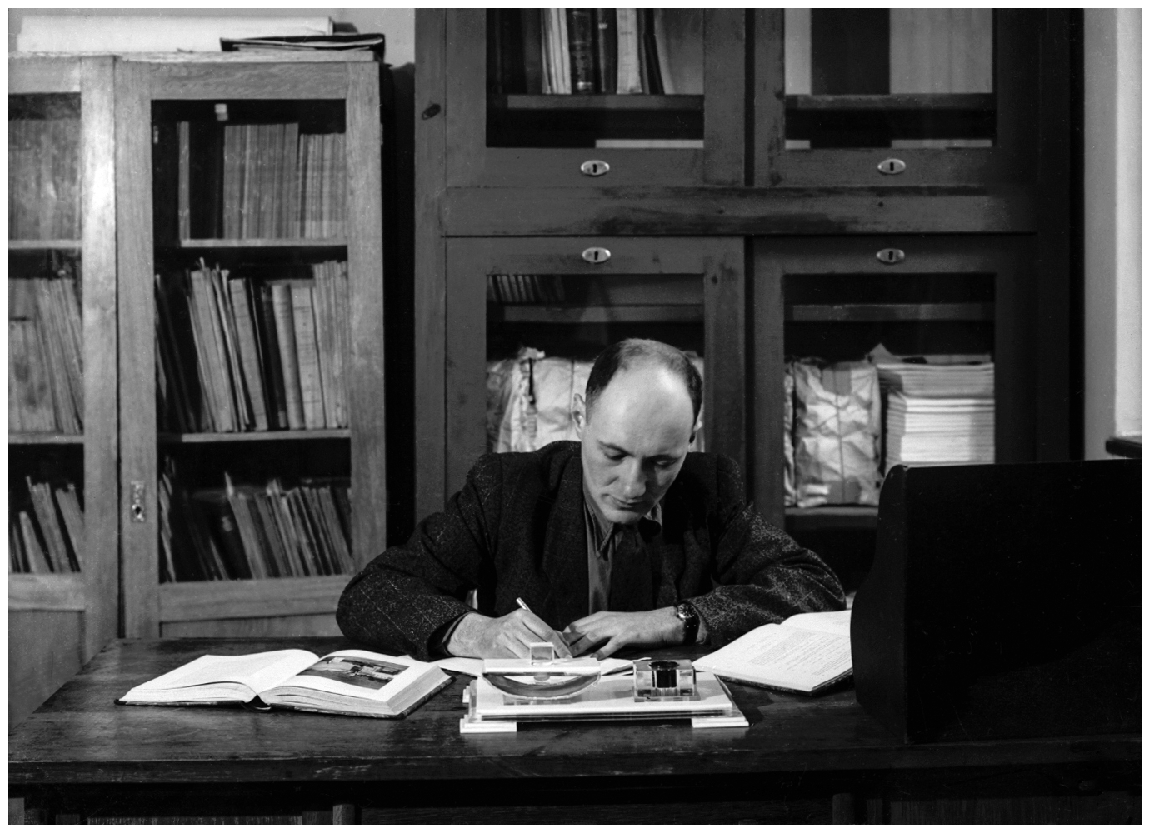}
\caption{Extract from the order on S.~Kaplan’s employment in the position of a head of the astrophysical sector and one of his first photographs of the Lviv period, which is stored in the photo archive of the Astronomical Observatory.}
\label{fig1}
\end{figure}

Samuil Kaplan was born on October 10, 1921 in Roslavl, Smolensk Region (Russian Federative Soviet Republic). From 1926 he lived in Leningrad with his parents. After graduating from high school in 1939, S. Kaplan entered the Leningrad Polytechnic Institute, but during the war he transferred to the Herzen Pedagogical Institute of Leningrad, from which he graduated in the spring of 1945 with a diploma in mathematics. From November 1939 to October 1945 he served in the active army, participated in the II World War, fought on the Leningrad front. After demobilization, he entered the graduate school of Leningrad University at the Department of Theoretical Astrophysics. Samuil Kaplan came to Lviv in June 1948 to work at the Lviv University as a head of the astrophysical sector of the Astronomical Observatory. This happens after he had successfully completed the graduate school at the Leningrad University and had defended his dissertation dedicated to ``Energy sources and the evolution of white dwarfs''. The referral certificate, which is stored in S.~Kaplan's personal file in the University Archives\footnote{ S.~Kaplan’s personal file, No 5583, the Archives of the Ivan Franko National University of Lviv.}, dates back to April 30, 1948, and the date of arrival indicated therein is June 1. He was awarded the degree of the Candidate of Physical and Mathematical Sciences on June 30, 1948. On June 10, S.~Kaplan filed an application to Prof.~Biliakevych, rector of the Lviv University, declaring his wish to work. He obtained the position of the head of the astrophysical sector with a salary of 2\,400 karbovanets\footnote{This is the currency at that time.} on July 1, 1948, as evidenced by an extract from the Order No 31 of the Astronomical Observatory issued on July 17 (Fig. \ref{fig1}). Due to the fact that only the half-time position was available, on July 1, S.~Kaplan was additionally hired (with 0.5 rate compensation) for a position of a senior lecturer at the Department of Experimental Physics with a salary of 1\,250 karbovanets per month (Order No 253, July 26). The combination of the positions of a researcher and a lecturer was not allowed at that time. Consequently, on September 1, he was transferred to the position of a full-time senior lecturer in the Department of Theoretical Physics (Order No 287, August 31).

In S.~Kaplan's file, there is the personal characteristic sent from the postgraduate school of the Leningrad University. It was sent on September 25, 1948, upon the request of vice-rector K.~M.~Leutskyi, and sent again on July 19, 1949, following the request of the director of the Observatory Oleksander Syrokomskyi. They state that during his postgraduate studies, S.~Kaplan made several scientific reports at the seminars demonstrating significant erudition in the field of internal structure of stars and predisposition to theoretical work, but as a person, he was characterized as too self-confident. Both characteristics mention his ``unpatriotic'' action, the essence of which was a overview of the works of foreign scientists in the theory of white dwarfs at the jubilee scientific session of the Faculty of Mathematics, held on the occasion of the 30th anniversary of the October Revolution, where he did not mention the works of soviet scientists in this area, and "the leading role" of the communist party of the Soviet Union. He was reprimanded for this, which, apparently, made it impossible for him to work at the Leningrad University at the times of the struggle against cosmopolitism.

After obtaining the full-time position of a head of the аstrophysical sector, S.~Kaplan was transferred to the Observatory, which became his main place of work (transfer application dates back to December 30, 1948); though, he also remained in a half-time position of a senior lecturer in the Department of Theoretical Physics, where he was affiliated until the end of the Lviv period of his activity.

\begin{wrapfigure}{i}{0.44\textwidth}
\includegraphics[width=0.44\textwidth]{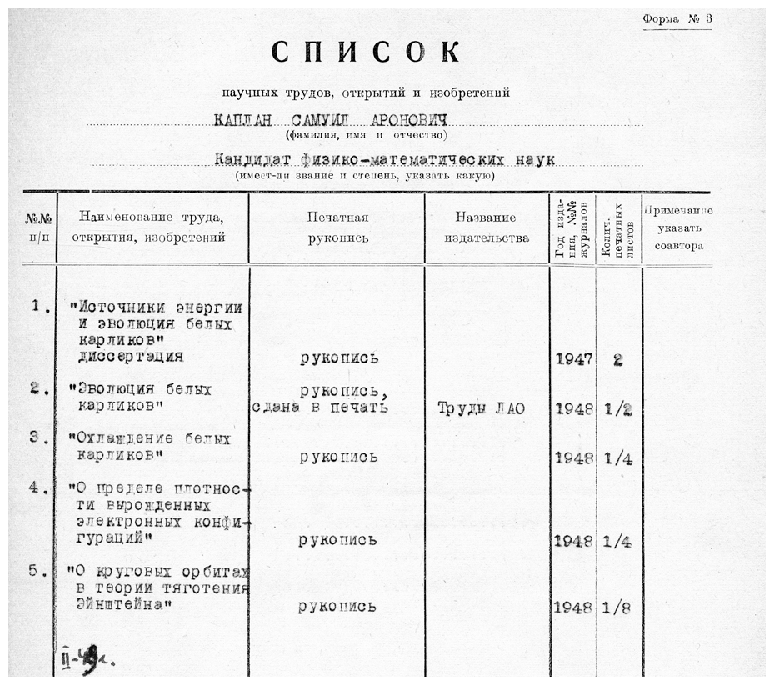}
\caption{The first list of S.\,Kaplan’s publications found in his personal file.}
\label{fig2}  
\end{wrapfigure}

At the Lviv University, Samuil Kaplan embarked on research in the field of theoretical astrophysics and had an opportunity to choose the relevant areas to study. It was here that he started working actively, as evidenced by the first list of his publications provided in his personal file (Fig. \ref{fig2}) dating back to February 1949 and the bibliography of S.~Kaplan’s works of the Lviv period, which includes 95 items (it may be found at the end of this paper). 

Giving lectures to students on electrodynamics, solid mechanics, thermodynamics and statistical physics, radiation theory, general relativity, theoretical astrophysics and the internal structure of stars contributed to his growth as a scientist, they gave him the opportunity to solve the most actual problems of theoretical astrophysics at that time, which made him a universal astrophysicist of the twentieth century. Volodymyr Pronyk, one of S.~Kaplan's students, recalls his teacher’s words in his memoirs \cite{Shklovski}, ``When you want to study a section of physics or astrophysics, deliver a course of lectures for students.'' Another S.~Kaplan's student, Oleksandr Lohvynenko, recalls that Samuil Kaplan sincerely shared his knowledge with students, ``each of his lectures was original; he created on the board, the presentation of the most difficult issues was digestible and clear, and the pace of the presentation made it possible to summarize and note everything'' \cite{Lohvynenko2011}.

In February 1950, a new director of the Astronomical Observatory, Volodymyr Stepanov, appointed S.~Kaplan the deputy director of scientific work (with 3\,200 karbovanets salary). He held this position until 1953, when Moris Eigenson took over as a director after V.~Stepanov's transfer to the Sternberg State Astronomical Institute (Moscow). In the same year, according to an excerpt from the protocol of the Higher Attestation Commission of the USSR Ministry of Higher Education (June 3), and the Order No 365 of the rector of the Lviv University issued on July 11, Samuil Kaplan’s academic title as a senior researcher specializing in astrophysics was approved. He was actively involved in scientific and organizational work. In 1950, S.~Kaplan took part in selecting a site for the construction of the suburban observation station in the village of Briukhovychi, and in 1952, he managed to obtain a resolution of the Council of Ministers of the USSR which allowed the construction. In the same year, he led an expedition to the Shieli district of the Kyzylorda region of the Kazakh SSR to observe a total solar eclipse, which occurred on February 25. S.~Kaplan supervised the development of the design of the AZT-14 telescope by the Leningrad Optical-Mechanical Association for the astrophysical sector of Lviv Observatory, which was later renamed as the Department of Variable Stars. In S.~Kaplan's personal file, we found a statement dated January 2, 1953, in which he requested to be sent to Moscow and Leningrad free of charge to monitor the process of the design of the 50-cm reflector and to work in libraries from January 12 to February 5. There are several similar statements in the researcher’s personal file. S.~Kaplan was often invited to participate in scientific conferences held in Moscow or Leningrad at the expense of the organizing committee. He regularly visited the Simeiz Astronomical Observatory in Crimea, where he befriended the Shajn and Pickelner families. After the first works published in the late 1940s on compact, very dense, degenerate stellar objects, in the early 1950s, S.~Kaplan began studying rarefied astrophysical objects, i.e. atmospheres of stars, gas nebulae, interstellar medium, and space plasma.  

S.~Kaplan's research and educational career took off quickly up. On September 1, 1952, he was transferred to the position of acting Associate Professor of Theoretical Physics Department, and by the Order of the rector No 514 issued on July 21, 1954, he was approved for this position based on the results of the competition. S.~Kaplan worked part-time as a senior researcher in the Observatory where he had a permanent workplace. However, on April 1, 1956, he was abruptly dismissed from the position of senior researcher due to the discovery of violations of a rule during an audit of the University's financial activities. The reason for this was the Order No 213 of the Minister of Higher Education of the USSR B.~A.~Koval issued on March 19, 1956. Despite this fact, on September 20, 1957, S.~Kaplan was appointed as a head of the optical observation station of the artificial earth satellites ``Lviv-1031'' established under the auspices of the Astronomical Observatory and subordinated to the Astronomical Council of the USSR\footnote{In the extract from the Order No 678 of Ivan Franko State University of Lviv, it is listed as the optical station of the Faculty of Physics.}. He was performing his duties until November 25, 1958, and after that, handed the station over to I.~Klymyshyn.  

\begin{wrapfigure}{i}{0.5\textwidth}
\includegraphics[width=0.49\textwidth]{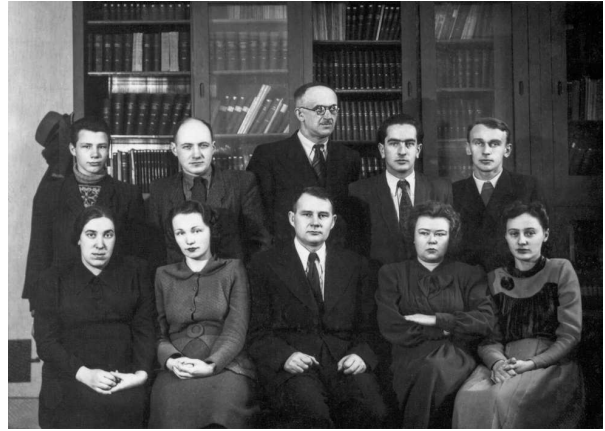}
\caption{The staff of the Astronomical Observatory in the early 1950's. From left to right: (sitting) T.\, Mandrykina, R.\,Teplytska, V.\,Stepanov, N.\,Yelenevska, (?); (standing) (?), S.\,Kaplan, Ya.\,Kapko, A.\,Kopystianskyi, H.\,Rodionov.}
\label{fig3}  
\end{wrapfigure}

In S.~Kaplan's personal file, there are copies of his characteristics issued for various purposes.  Interestingly, a record dating back to November 10, 1955, signed by rector Ye.~Lazarenko and appointed to be presented at the Moscow University on the occasion of the defense of the scientific degree of Doctor of Science, stated that during his work at the University S.~Kaplan ``proved to be a capable researcher and a good teacher. He has published more than 35 scientific papers and co-authored with K.~P.~Staniukovych and F.~A.~Baum a book ``Introduction to space gas dynamics''. He works a lot on training physicists and astrophysicists at the Lviv University, supervises the writing of dissertations and PhD theses ...'' It was also mentioned that S.~Kaplan had enjoyed a well-deserved authority among teachers and students, participated in the public life of the University, constantly increased the ``level of ideological and political awareness'', often gave popular lectures, and managed to complete his doctor of sciences thesis.

From yet another personnel record sheet, filled in by S.~Kaplan himself, we have learned that he defended his Doctor of Science thesis\footnote{The Soviet system of scientific degrees was two-level: the 1st scientific degree, candidate of sciences, is similar to a PhD, the 2nd scientific degree, doctor of science, is similar to a habilitated doctor in some countries. Each level was achieved by the applicant after a positive evaluation of the corresponding thesis and its public defence in the special councils.} on ``Methods of gas dynamics of the interstellar medium'' at the  M.~Lomonosov State University, in Moscow on March 1 1957, at the age of 35. The copy of the diploma states that the decision to award him a degree of the Doctor of Physical and Mathematical Sciences was made by the USSR Higher Attestation Commission on January 18, 1958, and the diploma МФМ No 000176 was issued on February 4. A few months earlier, on October 11, 1957, by the order of the rector of the University, S.~Kaplan was transferred to the position of acting Professor of Theoretical Physics Department, without waiting for an official granting of the diploma, and on May 19, 1958, S.~Kaplan was approved for this position as the one selected by competition. On November 18 of the same year, he was awarded the academic title of the Professor of Theoretical Physics Department.

\begin{figure}[!t]
\includegraphics[width=0.48\textwidth]{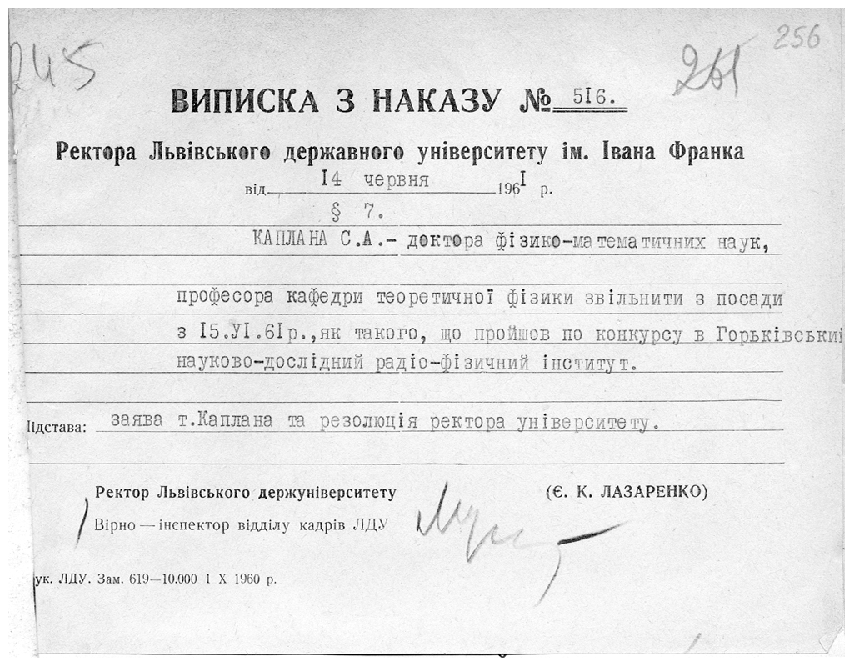}
\includegraphics[width=0.51\textwidth]{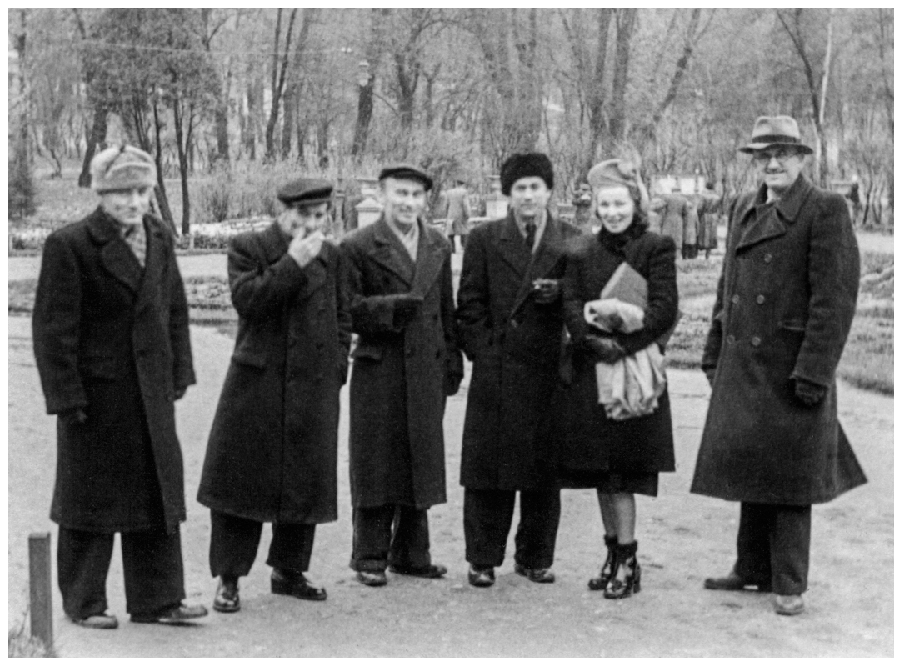}
\caption{Extract from the order on dismissal of S.~Kaplan and one of the photos of the late 1950s from the photo archive of the Astronomical Observatory, where S.~Kaplan is on the far left.}
\label{fig4}
\end{figure}

In 1959, there was a sudden change of the director of the Astronomical Observatory; the position was obtained by Yaroslav Kapko, Master of Astronomy and a member of the Communist party. It seems that S.~Kaplan realized that he as a Doctor of Science but not a member of a Communist party had no chance to lead a research team neither on the level of a department nor a research institute of the University. Therefore, he started to look for an opportunity to change his place of work and get an offer from the institutes closer to Moscow or Leningrad where his colleagues and co-authors worked back then. These thoughts appear based on the 1960 characteristics report which was supposed to be presented for the Pulkovo Observatory. However, it was obvious that the ``capital'' institutions were still closed for S.~Kaplan. Thus, he accepted the offer of M.~Grekhova, director of the Gorky\footnote{Now Nizhny Novgorod, regional center in central Russia} Research Institute of Radiophysics, and moved there in 1961, as evidenced by an extract from the Order No 516 of June 14 (Fig. \ref{fig4}) of rector of Lviv University. Obviously, the given hypothesis should be viewed as an attempt to comprehend why S.~Kaplan left Lviv, the city he loved, the University where he was comfortable working, had students and friends, and enjoyed the authority among lecturers and learners. After all, Lviv is a place where his son Jacob was born in 1956. A few of his colleagues from the Gorky Research Institute of Radiophysics\footnote{Pankrashkina N.G., Sitkova Z.P., Samuil Aronovich Kaplan, http://www.itmm.unn.ru/ob-institute/nemnogo-istorii/memorial/samuil-aronovich-kaplan/} shared a similar opinion in their memoirs saying that ``despite the fact that he was optimally equipped in terms of scientific life, he still complained about the lack of a laboratory or any analogue to a laboratory, which was autonomous and officially recognized''. 

The years S.~Kaplan spent working at the Gorky Institute of Radiophysics, and later the Gorky State University (part-time), were as fruitful as the years spent in Lviv. He worked there until the day of his tragic death\footnote{Returning from a business trip to Leningrad, he died at the Bologoye station, having fallen under a train.} on June 11, 1978. In 1980, to commemorate the researcher, the asteroid number 1987 discovered by Pelageya Shajn at the Simeiz Observatory (Crimea) in 1952, was named ``Kaplan''.  

In the following sections, we will discuss Samuil Kaplan’s works performed during the Lviv period in the light of the current state of those areas of research.

\section{Stability of circular orbits in Einstein's theory of gravitation}


In the bibliography of Samuil Kaplan's works of the Lviv period, the first item [1]\footnote{Such numbers refer to papers from the Bibliography of Samuil Kaplan in the Lviv period presented at the end of this paper.}, chronologically, is the short paper, slightly bigger than one page, published in ``The Journal of Experimental and Theoretical Physics''. It appeared as a letter to the editor, and the original title read as follows, ``On circular orbits in Einstein's theory of gravitation''. Even though at that time the author was only 28 years old, his work gained international recognition among scientists working in the field of relativity and astrophysics. However, this did not happen instantly, but about 25 years later, after the publication of a fundamental monograph on the theory of gravitation by famous authors C.~Misner, K.~Thorne and J.~Wheeler \cite{Misner}. In its extensive bibliography on p. 1238, it was stated, ``Kaplan~S.~A.~1949b. O krugovykh orbitakh v teorii tyagoteniya Einsteina (On circular orbits in Einstein’s theory of gravitation)''. In the text of the monograph, though, where on p. 662 the properties of geodetic orbits in the Schwarzschild field are presented using the results obtained by S.~Kaplan, there is no mention of him. The same omission persisted in the translation of the Misner, Thorne and Wheeler’s monograph into Russian, which had three volumes. S.~Kaplan's pioneering publication is mentioned in the second volume \cite{Misner2}, in the bibliographic list No 261 on p. 512 (unlike in the original, in translation the bibliography is presented without the paper’s titles).

In the famous book written by L.~Landau and E.~Lifshitz \cite{Landau-Lif} on p. 399, the following problem is formulated, ``For a particle in the collapsar field, find the radii of circular orbits (S.~A.~Kaplan, 1949)''. It also provides a solution to this problem and reproduces the results of S.~Kaplan's paper on the regions of existence of circular orbits and the determination of the values of the Schwarzschild radial coordinate when they are stable or unstable (unfortunately, the title of S.~Kaplan's publication and the journal where it was published are not specified).

One more fundamental monograph on the theory of black holes \cite{Chandra-sek} cannot be overlooked. In fact, the authors described the properties of circular geodetic orbits in the Schwarzschild field utilizing S.~Kaplan’s approach, without specifying his publication issued in 1949. 

That is, the approach to the analysis of geodetic orbits in the Schwarzschild field, introduced by S.~Kaplan, proved to be effective. It is based on the use of Hamilton-Jacobi equations of geodetic lines instead of the traditional system of ordinary differential equations. 

Undoubtedly, there is a certain aesthetic in S.~Kaplan's brief study. Yet, in the first paragraph, he noted that he considered his research in the broader context of studying the properties of real physical configurations. This area of researh is reflected in many of his subsequent publications.

The paper ``On circular orbits ...'' is well known among experts, as evidenced by the site Astrophysics data system (https://ui.adsabs.harvard.edu/abs/1949ZhETF..19..951K/citations): it points out 32 citations of this paper for the period 1969-2016. In the absence of English translation, it is not available to a wider range of experts in the world. To eliminate this injustice, we translated the given paper into English and published it in the preprint library at arXiv: 2201.07971, which can also be found in the database of astrophysical publications https://ui.adsabs.harvard.edu/.

\section{Stability and cooling of white dwarfs}

The discovery of white dwarfs \cite{Adams_1915} has created a problem of sources of energy and the stability of stars, in which thermonuclear reactions are either absent or occur only in the outer layers with low intensity, and therefore, play a secondary role. According to R.~Fowler's idea \cite{Fowler_1926} the stability of white dwarfs is ensured by a quantum effect, i.e. the degeneracy of the electron subsystem, when $k_B T$ is much less than the energy of the electron on the Fermi surface. S.~Chandrasekhar generalized this idea for the case of high densities at which the electron subsystem is relativistic, and constructed the theory of cold dwarfs at $T=0$ К based on the equation of hydrostatic equilibrium~\cite{Chandrasekhar_1931}.

S.~Chandrasekhar's model embraces two components, namely a completely degenerate relativistic ideal electron subsystem and a subsystem of static nuclei, which is considered as a continuous classical medium. The equilibrium between the electron gas pressure and the gravitational compression created by the nuclear subsystem ensures the stability of the dwarf, and its characteristics (mass, radius, density distribution) are functions of two dimensionless parameters, i.e. the relativistic parameter $x_0 =\hbar (m_0 c)^{-1} (3\pi^2 n (0))^{1/3}$ (where $n (0)$ is number density of electrons in the center of the star) and the parameter of chemical composition $\mu_e =\langle A/z\rangle \approx 2.0$, where $A$ is mass number, $z$ is nucleus charge. The main conclusions reached by S.~Chandrasekhar are the restriction on the maximum mass of the dwarf (Chandrasekhar limit) and the specific mass-radius relation. From the modern point of view, S.~Chandrasekhar's model is very simplistic as it does not take into account such factors of structure formation as interparticle interactions, thermal effects, axial rotation, magnetic fields and the effects of the general theory of relativity. This is the basic model that explains the existence and stability of cold (black) dwarfs. It cannot explain the radiation of dwarfs, as well as the basic details of the distribution of dwarfs in the mass-radius plane \cite{Vavrukh_2018b}, which are known due to modern observations.

The issue of energy sources of white dwarfs was actively discussed from 1939 to 1952 in a number of publications. A.~Marshak~\cite{Marshak_1940} believed that the luminosity of dwarfs is caused by gravitational compression but did not study this mechanism. E.~Schatzman~\cite{Schatzman_1945,Schatzman_1947} adhered to the idea of thermonuclear reactions in the surface layers of dwarfs, which led to unphysical conclusions.

It was S.~Kaplan who offered a well-grounded idea from the physical point of view; according to it, dwarfs emit thermal energy accumulated in the past due to thermonuclear reactions. The heat reserve of the dwarf is distributed between its two subsystems, and it can be estimated using the equation of state. In the case of a massive hot dwarf, when the nuclear subsystem can be considered as an ideal classical gas and the electron subsystem has a small deviation from the absolute degeneracy, the equation of state takes the following form \cite{Vavrukh_2018b}:
\begin{equation}
\label{eq_01k}
P (r) = \frac{\pi m^4_e c^5}{3 h^3}\left\{ {\cal F}(x)+ \frac{4\pi^2}{3} T^2_* (r) \frac{x (2+x^2)}{\sqrt{1+x^2}}+\frac{8}{z} x^3 T_* (r) +\cdots\right\}.
\end{equation}
Here  $x \equiv x (r)$ is the local value of the relativistic parameter on the sphere of radius $r$; ${\cal F}(x)$ is the contribution of the electron subsystem at $T=0$ К;  $T_* (r) \equiv k_B T (r)/m_e c^2$; contribution, proportional to $T^2_* (r)$, generated by the deviation of the electron subsystem from absolute degeneracy; the last contribution in the curly braces of the formula (\ref{eq_01k}) is the contribution of the nuclear subsystem in the accepted approximation. In the case of a massive dwarf, when $\langle x (r)\rangle \:> 1$, and $\langle T_* (r)\rangle \ll 1$, the second term in the curly braces can be neglected. It corresponds to S.~Kaplan’s model, in which thermal energy is mainly concentrated in the nuclear subsystem, the average kinetic energy value of which is equal to~[2,4]
\begin{equation}
\label{eq_02k}
W_T \approx \frac32 \:\frac{k_B}{m_u \mu_n} \int \rho(r) T (r) d{\bf r} \cong\frac32\: \frac{k_B}{m_u \mu_n} M T,
\end{equation}
where
$m_u$ is atomic mass unit, $\mu_n$ is dimensionless molecular mass of the nucleus $(\mu_n = 2z)$, $T\equiv \langle T (r) \rangle$. Luminosity is determined based on the relation (\ref{eq_02k})
\begin{equation}
\label{eq_03k}
L \approx - \frac{dW_T}{dt} \approx-\frac32 \:\frac{k_B}{m_u\mu_n}\:M\:\frac{dT}{dt},
\end{equation}
where $t$ is time. On the other hand, the value of the temperature in the inner part of the star (which is almost isothermal due to the electron nature of thermal conductivity) can be estimated by E.~Schatzman’s formula~\cite{Schatzman_1947}
\begin{equation}
\label{eq_04k}
T = T_0 \:\left\{\frac{L/L_{\odot}}{M/M_{\odot}}\right\}^{2/7}; ~~~~T_0 = 6.16\cdot 10^7 \mbox{\rm K}.
\end{equation}
From the expressions (\ref{eq_03k}) and (\ref{eq_04k}) we obtain a differential equation
\begin{eqnarray}
\frac{dT}{dt}=- \frac32 \:\frac{m_u\mu_n L_{\odot}}{k_B M_{\odot}}\left(\frac{T}{T_0}\right)^{7/2},\nonumber
\end{eqnarray}
the solution of which determines the change in temperature over time
\begin{eqnarray}
T^{-5/2} (t)- T^{-5/2}(t_0) = (t-t_0)\:\frac53\:\frac{m_u\mu_n}{k_B}\:\frac{L_{\odot}}{M_{\odot}}\:
T^{-7/2}_0, \nonumber
\end{eqnarray}
where $T (t_0)$ is temperature at the initial moment $t_0$. Given that $T (t_0) \gg T (t)$, we see that
$T (t) \sim (t-t_0)^{-2/5}$. Taking into account the expression (\ref{eq_04k}), we can calculate the cooling time
\begin{eqnarray}
\tau \equiv t-t_0 = \tau_0\:\left(\frac{M/M_{\odot}}{L/L_{\odot}}\right)^{5/7}; \quad
\tau_0 =\frac35\:\frac{k_B T_0}{m_u\mu_n}\:\frac{M_{\odot}}{L_{\odot}}\approx \frac{5}{\mu_n}\:10^7 \:\mbox{years}.
\nonumber
\end{eqnarray}
Provided that $L /L_{\odot} \approx 10^{-3}$,  cooling time $\tau \approx 5/\mu_n\:10^9$ years. Such a large cooling time is due to the small temperature gradient inside the dwarf, as well as a significant reserve of thermal energy of the nuclear subsystem.

For example, consider the dwarf Sirius B, which has an effective temperature $T_{eff}\approx 25\cdot 10^3 $ К, mass $1.02\: M_{\odot}$, luminosity $2.5\cdot 10^{-2}\: L_{\odot}$, radius 5668 km, average density $\langle \rho\rangle =2.5\cdot 10^6 ~\mbox{g/cm}^3$, relativistic parameter in the center $x_0 \cong 2.4$. The reserve of the thermal energy in its nuclear subsystem is close to
\begin{eqnarray}
\frac{\pi m^4_0 c^5}{3h^3}\:\frac{8}{z}\int\limits_{V} x^3 (r)\:T_* (r)\:d{\bf r}\approx \frac{5}{z} \cdot10^{48} \:\mbox{erg}.\nonumber
\end{eqnarray}
At $T_* \approx 10^{-2}$ it will suffice for the time exceeding $10^8$ years. It should be noted that the energy of the black dwarf with the parameters of the star Sirius B (kinetic energy of the electron subsystem + gravitational energy of the nuclear subsystem) is close to ($-10^{50}$ erg). Thus, thermal energy constitutes a few hundredths of the total energy and, consequently, has little influence on the structure and stability of the dwarf, which confirms the adequacy of S.~Chandrasekhar's model.

The monograph \cite{Shapiro_1983}, which is the fullest review of the studies of white dwarfs over the past century, does not refer to the work of S.~Kaplan, but there is a reference to the work of L.~Mestel~\cite{Mestel_1952}, published 3 years later, which repeated S.~Kaplan’s calculations. However, S.~Kaplan's priority is unquestionable. It is interesting to see how quickly he formulated his position on the energy sources of white dwarfs. If in the work [2] S.~Kaplan (apparently under the influence of A.~Marshak's idea), wrote that the radiation of dwarfs can be viewed as a result of gravitational compression, in the work [4] the idea of their cooling is clearly formulated. Indeed, when the dwarf is cooled, there is a reduction in its size, which leads to some energy release, but this mechanism makes a small contribution, it is a consequence of cooling and somewhat prolongs the cooling time. The primary cause is the cooling of the dwarf, which is a consequence of the existence of a temperature gradient along the radius of the star.

The discovery of white dwarfs raised the general question of the possibility of existence of other forms of matter at high density. L.~Landau suggested the hypothesis of the existence of neutron stars \cite{Landau_1932}, which would be analogous to white dwarfs. In this regard, the question of the limits of stability of white dwarfs had become extremely relevant. S.~Chandrasekhar's model does not restrict the density of matter in white dwarfs; mass is a monotonically increasing function of the relativistic parameter $x_0$, and radius is a monotonically decreasing one. As shown by E.~Schatzman~\cite{Schatzman_1947}, at $x_0 \gg 1$
\begin{eqnarray}
M (x_0) \:=\: M_3\:\left\{1 - \frac{a_2}{x^2_0} +\frac{a^4}{x^4_0}\:+\:\cdots\right\},
\quad  
\mbox{\rm where}
\quad
M_3 \:\cong\:\frac{5.76}{\mu^2_e}\:M_{\odot}\approx 1.45 M_{\odot} \nonumber
\end{eqnarray}
is the mass of the Emden polytrope with the index $n=3$. Since white dwarfs were considered to be the predecessors of neutron stars, the task of studying white dwarfs at high densities (when the effects of the general theory of relativity are manifested) had become urgent; it required going beyond the Newtonian approximation on which S.~Chandrasekhar's theory was based. In order to do this, S.~Kaplan used the so-called Oppenheimer-Volkoff equation \cite{Shapiro_1983}, which was intended to describe the then hypothetical neutron stars. In the works [3,58], solutions of this equation were obtained for white dwarfs considering the effects of the general theory of relativity with an accuracy of $c^{-2}$ according to the perturbation theory. In this approximation, the mass of the dwarf can be asymptotically represented as
\begin{equation}
\label{eq_11k}
M \:=\: \frac{M_0}{\mu^2_e}\:\{M_0 (x_0)\:-\:\gamma (x_0)\:M_1 (x_0)\:+\:\cdots \},
\end{equation}
where
$M_0 \approx 2.89\: M_{\odot}$, $\mu_e \approx 2$, $M_0 (x_0)$ is dimensionless mass of the dwarf in S.~Chandrasekhar’s model, which is a monotonically increasing function of the parameter $x_0$, which goes to the value 2.01824 at $x_0\gg 1$, $M_1 (x_0)$ is a positive function that weakly depends on $x_0$ ($4.09 \leq M_1 (x_0)\leq 4.11$ within $20\leq x_0 \leq 30$),
\begin{eqnarray}
\gamma (x_0)=\varepsilon (x_0)\:\frac{m_e}{m_u\mu_e}\:\approx \frac{x_0}{4}\:10^{-3}.\nonumber
\end{eqnarray}
The function (\ref{eq_11k}) has a maximum in the vicinity $x_0\approx 26$ when $\frac{m_e}{m_u\mu_e}\:M_1 (x_0)=\frac{d}{dx_0}\:M_0 (x_0)$. This value of the relativistic parameter at $\mu_e =2$ corresponds to the central density $\rho_c^{\mbox{\scriptsize cr}}\approx 2.1\cdot 10^{10}\: \mbox{g/cm}^3$, and critical mass $M_{\mbox{\scriptsize cr}}\approx 0.96\:M_3$. Taking into account the effects of the general theory of relativity places a limit on the mass of the dwarf in S.~Chandrasekhar’s model. Considering the Coulomb interparticle interactions and axial rotation causes an additional reduction in the critical mass of the dwarf \cite{Vavrukh_2018b} by reducing $M_0 (x_0)$, which depends on both the charge of the nucleus $z$ and the angular velocity of rotation. However, S.~Kaplan in his works was the first one to prove that there is a limit on the densities of electron configurations due to the effects of the general theory of relativity. Although neutron stars were discovered 18 years after the publication of S.~Kaplan’s works, and white dwarfs with masses corresponding to $x_0 \approx 20-25$ were not recorded, S.~Kaplan’s research, devoted to the stability of electronic configurations, played a fundamental role in the development of the theory of white dwarfs.

\section{Properties of shock wave motion in inhomogeneous media}

Among the unique in volume and depth S. Kaplan's scientific works of the Lviv period, the study of magnetohydrodynamic processes in space plasma occupies an important place. In the 1950s, physics of the interstellar medium became the front line of astrophysical research laying the basis of the cosmogonic theory of the star formation and evolution, where magnetohydrodynamic processes determine the physical aspects of the collapse of protostellar molecular clouds, the formation and subsequent evolution of stars, the final stages of their existence, and the evolution of the Supernova remnants. A key role in the dynamics of the interstellar medium is played by shock waves, which provide rapid dissipation of plasma kinetic energy accompanied by heating of plasma to high (keV) temperatures and acceleration of cosmic rays to the energy of about 10$^{17}$ eV. Cosmic rays, in turn, become effective sources of non-thermal synchrotron (radio -- X-ray range) and inverse Compton (X-ray -- gamma-ray range) radiation. In the 1950s, the radio astronomical window had already been effectively used for the studies of non-thermal radiation from the objects of the interstellar medium. Most of S.~Kaplan’s publications of the Lviv period are devoted to the problems of magnetohydrodynamics of the interstellar medium [10-11,13-24,26-39,41-42,46-53,55-57,59-62,64-65,67-83,85-94]. The study of magnetohydrodynamic processes associated with shock waves began in 1955 [31]. Since 1959, the given research papers were co-authored by Ivan Klymyshyn [60-62,90-91], who continued to study the physical processes in the shells of the stars and the interstellar medium, which are related to the generation and propagation of shock waves. The results of I.~Klymyshyn's research are summarized, in particular, in well-known monographs
\cite{Klimishin1972,Klimishin1984}.

As a result of the ability of shock waves to heat cosmic plasma and transfer part of their energy (about 10\%) to newly accelerated cosmic rays, their presence in astrophysical objects is manifested by thermal radiation of the heated plasma (X-ray thermal radiation of the Supernova remnants, etc.), and non-thermal radiation of the accelerated cosmic rays (non-thermal radio radiation of the Supernova remnants, high-energy (GeV-TeV range) gamma-ray emission of relativistic jets of active galactic nuclei, etc.). At the same time, the nonlinearity of the equations of magnetohydrodynamics which describe the flows of plasma with shock waves significantly complicates the analytical description of the dynamics of shock waves in the astrophysical objects. The only complete analytical description of plasma flows with strong shock waves is Sedov's solution of the problem of self-similar plasma flows with shock after a point-like explosion in the uniform (with constant density) and in the non-uniform medium with the power-law density distribution. These solutions are widely used in astrophysical modeling of the processes associated with rapid localized energy release, ranging from solar flares to the evolution of the Supernova remnants. However, in most astrophysical applications, shock waves propagate in non-uniform media with the loss of self-similarity and can only be modeled by numerical or approximate analytical methods.

In the second half of the twentieth century, numerical simulations were severely limited by the computer capabilities, thus, much attention was paid to the development of approximate analytical methods for describing the motion of shock waves in the inhomogeneous astrophysical environments. I.~Klymyshyn, in particular, was engaged in the development and improvement of the approximate analytical methods; the results he obtained are presented in \cite{Klimishin1972}. 
 
In the 1950s, the central task was magnetohydrodynamic description of the plasma flows caused by the energy release in solar flares when the shock wave profile is significantly non-spherical and extends in the direction opposite to the density gradient in the solar corona and the solar wind. Notably, S.~Kaplan utilized the approximate analytical methods to describe the dependence of the speed of propagation of the shock wave front (caused by the solar chromospheric flash on October 4, 1965) from the distance to the Sun. The Type II solar flare was observed at frequencies of 0.21 MHz and 2 MHz and its temporal evolution could be explained in the framework of non-thermal radiation of the shock wave. At that time, the Brinkley-Kirkwood and Chisnell-Whitham analytical methods \cite{Klimishin1972} were the most developed ones to describe the dynamics of shock waves. Samuil Kaplan was the first to note that in the case of strong shock waves both methods can help find the first integral of the equations of motion, which allows us to write the law of motion of a strong shock wave in the inhomogeneous medium with variable density $\rho (r)$ in the following form:
  \begin{eqnarray}
  D(r)=const\cdot\rho^a \cdot S^b
  \nonumber  
  \end{eqnarray}
where $r$ is the distance from the place of the shock wave generation (energy release), $S\propto r^N$ is a geometric divergence factor, $N$=0,1,2 for plane, cylindrical and spherical shock wave, respectively, $a=a(\gamma)$ and $b=b(\gamma)$ -- parameters that are weakly dependent on the plasma adiabatic index ($\gamma$=5/3 for an ideal gas). Since in both methods $a\approx b \approx -0.25$, S.~Kaplan suggested an approximation
\begin{equation}
	D(r)=const\cdot(\rho\cdot S)^{-1/4} 
	\label{2}  
\end{equation}
as a universal law of motion of a strong shock wave in an arbitrarily inhomogeneous medium with variable density $\rho(r)$ \cite{Kaplan1967}.
The approximation (\ref{2}) qualitatively corresponds to the numerical simulation of the motion of shock waves in the environments with a rapid drop in density where
$m^*(r)=- d\ln\rho(r)/d\ln(r)> N+1$, but is not applicable to media with slowly descrealing, constant and increasing density where $m^*(r) \leq N+1$. Therefore, in I.~Klymyshyn and B.~Gnatyk’s work \cite{Kli-Gna1981} on the basis of numerical modeling of strong shock waves in the environments with different types of density change and on the basis of analysis of approximate self-similar solutions on the regularities of strong shock waves in the environments with exponential and polytropic density distributions, a more accurate approximation formula for the motion of a strong shock wave in an arbitrarily inhomogeneous medium is offered. The initial stage of the motion of a strong adiabatic shock wave in realistic astrophysical media ($m^*(0)=0$) is always a Sedov-type deceleration: 
\begin{equation}
	D(r)=D_{dec}(r)=\frac2{3+N}\cdot \left(\frac{E}{\alpha_A\rho(r)} \right)^{1/2}r^{-(N+1)/2}, m^*(r)\leq N+1 
	\label{3} 
\end{equation}
If the condition $m^*(r)\leq N+1$ is satisfied on the whole interval $r$, the shock wave moves with deceleration all the time in the areas with $m^*(r)< N+1$ or at a constant speed in the areas with $m^*(r)= N+1$. If, starting from some distance $r_1$, the negative density gradient is less than the critical $m^*(r)> N+1$, the shock wave accelerates according to the law
  \begin{equation}
  	D(r)=D_{acc}(r,r_1)=D_{dec}(r_1)\cdot \left(\frac{\rho(r_1)r_1^{N+1}}{\rho(r)r^{N+1}} \right)^{1/5}, m^*(r)> N+1. 
  	\label{4} 
  \end{equation} 
Here $E$ is the energy of the point explosion that generates the shock wave, and $\alpha_A$ is the self-similar constant \cite{Kli-Gna1981,Klimishin1984}.

The equation of the trajectory of the shock wave can be derived from the following expression:
\begin{eqnarray}
t=\int_0^r\frac{dr'}{D(r')}.
\nonumber
\end{eqnarray} 
The aforementioned approximation gives the results close to approximate self-similar solutions in the field of their applicability and coincides with Sedov's self-similar solution in the case of homogeneous medium. The possibility of applying the approximation formula to describe the evolution of a shock wave in the general case of two- and three-dimensional motions in the absence of any symmetry appears to be important. In this case, the calculation of the evolution of the shock wave profile is carried out within the sector approximation, i.e., the three-dimensional region is divided into the required number of sectors and for each of them the trajectory equation is integrated.

The approximation formula (\ref{3})-(\ref{4}) in \cite{Gnatyk1985} was generalized to an arbitrarily relativistic case
\begin{eqnarray}
	\Gamma(r)\cdot\beta(r)=const\cdot(\rho(r)\cdot r^{N+1})^{-k},\nonumber
\end{eqnarray}
where $\beta=D/c$ is the speed of the shock wave, normalized to the speed of light $c$, $\Gamma=(1-\beta^2)^{-1/2}$ -- its Lorentz factor, $k=1/2$ and $k=1/5$ respectively, for the shock wave that decelerates ($m^*(r)\leq N+1$) and accelerates ($m^*(r) > N+1$).

Due to their high accuracy, the proposed approximations remain useful even when there is an opportunity to perform advanced computer simulations of the dynamics of shock waves in the inhomogeneous media.

\section{Gas-magnetic turbulence}

Samuil Kaplan's interest in the properties of shock waves and magnetic gas dynamics logically extended to another class of problems that had just started to be studied at the time, namely the description of the interstellar turbulence. The scientist devoted a number of his works [11,16,19,22,26-29,71,81,83], written in 1952-1958, to this topic. In these papers, S.~Kaplan has developed the theory of gas-magnetic turbulence (mainly works [26,27]) and methods for obtaining its characteristics from astronomical observations. The studies of plasma flows with shock waves, magnetic fields and turbulence resulted in the book [42], which consistently described and summarized the results obtained by S.~Kaplan in Lviv. It should be noted that after leaving Lviv, he continued to work on problems related to turbulence (see, for example, \cite{kaplan1972}).

At that time, the theory of turbulence in an incompressible fluid had already been well developed; there were the world-known studies published by A.~Kolmogorov, W.~Heisenberg, S.~Chandrasekhar, and others. A significant novelty of the results obtained by S.~Kaplan lies in the fact that he proposed the equations and analyzed the solutions which describe ``gas-magnetic'' turbulence, i. e. in the flows with shock waves and the consideration of gas compressibility is necessary, and a turbulent magnetic field is present. 

S.~Kaplan generalized the Heisenberg equation \cite{heis1948} for nonmagnetic turbulence in an incompressible fluid to the case of gas-magnetic turbulence. The equations describe the spectral functions of the kinetic $F(k)$ and magnetic $G(k)$ energies, which are the expansions of the corresponding values in the Fourier integral with the wavenumbers $k=2\pi/l$, where $l$ is the spatial scale of the corresponding turbulence harmonic. $F(k)dk$ and $G(k)dk$ are the densities of the kinetic and magnetic energies contained in vortices with wavenumbers from $k$ to $k+dk$, i.e., 
$$
F(k)=\frac{1}{2}\frac{dv_k^2}{dk},\qquad G(k)=\frac{d}{dk}\frac{H_k^2}{8\pi\rho},
$$
where $\rho$ is the gas density, $v_k$ and $H_k$ is the gas velocity and magnetic field strength in the vortices with scales $l=2\pi/k$. The Fourier decomposition of energies assumes isotropic homogeneous turbulence (otherwise the corresponding components of the correlation tensors should be decomposed). In the general nonstationary case, these equations describe the dissipation of the kinetic and magnetic energies in vortices of the corresponding scale per unit time (terms on the left) and have the following form:
\begin{equation}
-\frac{\partial}{\partial t}\int\limits_{0}^{k}Fdk=2\left(\nu+\kappa_\mathrm{f}y\right)
\int\limits_{0}^{k}Fk^2dk+2\int\limits_{0}^{k}\left(Fk^3\right)^{1/2}\left(\zeta_\mathrm{f}F+\mu G\right)dk,
\label{turb02}
\end{equation}
\begin{equation}
	-\frac{\partial}{\partial t}\int\limits_{0}^{k}Gdk=2\left(\lambda+\kappa_\mathrm{g}y\right)
	\int\limits_{0}^{k}Gk^2dk-2\int\limits_{0}^{k}\left(Fk^3\right)^{1/2}\left(\zeta_\mathrm{g}G+\mu G\right)dk.
\label{turb03}
\end{equation}
Here $\nu$ is the viscosity coefficient, $\lambda=1/(4\pi\sigma)$, $\sigma$ the electrical conductivity, $\kappa y$ is responsible for the turbulent viscosity or electrical conductivity, $\zeta$ and $\mu$ are dimensionless parameters close to the unity, indices 'f' and 'g' correspond to the kinetic and magnetic energies,
$$
y=\int\limits_{k}^{\infty}\sqrt{\frac{F}{k^3}}dk.
$$ 
The first terms on the right (proportional to $2\nu$ and $2\lambda$) describe the dissipation of energies on scales $>l$ into thermal energy due to viscosity or electrical resistance. The second terms (proportional to $2\kappa y$) describe the flow of the kinetic and magnetic energies from larger to smaller scales of turbulence. The third terms on the right (proportional to $2\zeta$) appear due to the presence of shock waves and represent the direct transition of the kinetic energy into the thermal energy and growth (hence the minus sign in the second equation before this term) of the magnetic energy in shock waves. The last terms in the equations (proportional to $2\mu$) with different signs describe the transformation of the kinetic energy into magnetic. 

S.~Kaplan solved the equation (\ref{turb02})-(\ref{turb03}) both in the stationary case and in the nonstationary case. In the latter case, he noted that the assumptions made during the solution were not sufficiently argued, so he did not provide solutions, but only analyzed them qualitatively. Regarding the stationary limit, in a certain approximation it appears that
\begin{equation}
F(k)=G(k)\propto k^{-\alpha}, \quad \alpha=\frac{5}{3}+\frac{32}{27}\frac{\zeta_\mathrm{f}-\zeta_\mathrm{g}}{\kappa_\mathrm{f}+\kappa_\mathrm{g}}+...
\label{turb01}
\end{equation}
that is, in a stationary gas-magnetic system there is an equality of total kinetic and magnetic energies, and they are evenly distributed among the vortices of different scales. In the absence of shock waves $\zeta_\mathrm{f}=\zeta_\mathrm{g}=0$, a well-known classical Kolmogorov spectrum $F(k)\propto k^{-5/3}$ is obtained. The presence of shock waves, i.e., the second term in the expression (\ref{turb01}) for $\alpha$, determines  the steeper turbulence spectrum comparing to the spectrum in an incompressible fluid, because the probability of occurrence and thus the dissipation of energy in small vortices is higher when shock wave is passing through the medium. 

In order to compare the theory with the observations, one needs to calculate the correlation functions and compare them to the velocities and strengths of the magnetic field at certain distances in the interstellar medium. Based on his solutions, S.~Kaplan deduced [42, p. 172] that the following should be expected from the theory: 
$$
\overline{(v_1-v_2)^2}\approx (\varepsilon r)^{2/3}(r/r_o)^\beta,\quad \beta=\frac{32}{27}\frac{\zeta_\mathrm{f}-\zeta_\mathrm{g}}{\kappa_\mathrm{f}+\kappa_\mathrm{g}},
$$
where $\varepsilon$ is the energy of sources of the interstellar turbulence, $r_o$ is its characteristic spatial scale. From the analysis of velocities of the interstellar clouds [29], S.~Kaplan derived that on scales smaller than $r_o=80$ pc, the following relation holds
$$
\overline{(v_1-v_2)^2}\approx (4\cdot 10^{-4}\ r)^{2/3} (r/80\ \mathrm{pc})^{0.05}.
$$
Such agreement with the observational data is impressive, but the researcher warned that due to the observed effects of selection (only sufficiently dense and slow clouds were observed), estimates $\varepsilon\approx 4\cdot 10^{-4}\ \mathrm{erg/cm^3}$ and $\beta\approx 0.05$  should not be considered very reliable.

\section{Acceleration of the cosmic rays}

During his stay in Lviv, S.~Kaplan published in 1953-1957 the four papers on acceleration of the cosmic rays [20,30,33,37]. Two of these publications were published in ``The Circular of the Astronomical Observatory of the Lviv University''. The latter is a generalization of the previous works in which the author's approach was gradually developed or described with varying degree of detail. 

It is worth noting that S.~Kaplan was quick to respond to E.~Fermi's pioneering work on the subject, which appeared in 1949 and 1954 \cite{fermi1949,fermi1954}. Cosmic rays have been known since 1912, but due to the complexity of experiments, progress in this area had been slow. For example, it took about ten years to understand and prove that cosmic rays are not really radiation but charged particles; another 10 years was needed to reveal that these are mostly positively charged particles (protons). E.~Fermi's interest in cosmic rays at that time was mainly due to the development of radio astronomy. Therefore, to get an idea of the state of development of science, it must be mentioned that the classic paper by Ginzburg and Syrovatsky on the theory of synchrotron radiation was published in 1965 \cite{gs1965ru,gs1965en}.

It must be mentioned that in his first work, E.~Fermi described the acceleration mechanism, which is now called the ``second order Fermi acceleration'', when the increase in the momentum of a relativistic particle $\Delta p$ is proportional to the second order of the average velocity of scattering centers $\Delta p\propto u^2$. In this process, shock waves are not involved in accelerating the cosmic rays. In the second work, E.~Fermi suggested the idea of recurrent acceleration in the presence of convergent plasma flows, when particles are successively reflected from one and the other ``mirror'', gaining energy regularly. The main for the present time ``Fermi mechanism of the first kind'' (when $\Delta p\propto u$), i.e., diffusion acceleration on shocks, was based on the similar idea of recurrence of the elementary processes, would be introduced later. Both S.~Kaplan and other authors at that time considered shock waves only as particle injectors, i.e., pre-accelerators to low energies (for example, during Supernovae explosions), after which the second-order acceleration can operate (e.g., \cite[p.~159]{gsbook1963}).

In his works, S.~Kaplan considered two mechanisms of acceleration in one approach: ``inductive'' (traditional acceleration of the electron in an electric field, which occurs, for instance, because of the changes in magnetic field) and ``statistical'' (the mechanism introduced by E.~Fermi), which also describes acceleration through the electric field $E$ which appears due to the motion of scattering centers with magnetic field $B$, i.e., $\mathbf{E}=[\mathbf{u}\mathbf{B}]/c$. Using such an approach, S.~Kaplan concluded that the average energy gain for particles is $\Delta \varepsilon/\varepsilon\propto u^2$, which is in agreement with E.~Fermi’s result. 

A significant novelty of S.~Kaplan's work, which was already present in the paper published in 1953, is that he considered how acceleration is affected by the turbulent properties of scattering centers. Namely, he calculated the diffusion coefficient of particles in the energy space through the turbulence spectrum $F(k,t)$
$$
D(\varepsilon,t)\approx \frac{p^2}{2\pi}\int\limits_{0}^{\infty}F(k,t)kdk,
$$
where $k$ is a wavenumber. S.~Kaplan wrote an interesting and important series of works on the theory of turbulence, which developed significantly in the Lviv period. Notably, he applied these results to the description of acceleration. 

In the aforementioned works, the author formulated the kinetic equation for the distribution function $N(\varepsilon,t)$ of particles with the term responsible for the Fermi acceleration (it contains $D(\varepsilon,t)$): 
$$
\frac{1}{u}\frac{\partial N(\varepsilon,t)}{\partial t}
=\frac{\partial^2}{\partial\varepsilon^2}\left[D(\varepsilon,t)N(\varepsilon,t)\right]
-\frac{\partial}{\partial\varepsilon}\left[\frac{d\varepsilon}{dx}N(\varepsilon,t)\right]+\frac{1}{u}\nu(E,t),
$$
He also took into account the terms responsible for different types of energy losses $d\varepsilon/dx$ (on radiation, nuclear reactions, ionization, etc.) and injection $\nu(E,t)$. In addition, S.~Kaplan claimed that the injection was provided by shock waves in which pre-acceleration was the result of energy dissipation of turbulent magnetic field movements. Thus, he wrote $\nu(\varepsilon,t)$ also through the turbulence spectrum $F(k,t)$. 
In his description, S.~Kaplan did not consider spatial diffusion (now the kinetic equation for acceleration on shock waves is of the diffusion type) arguing that, for the Fermi mechanism of the second order, it was sufficient to consider isotropic homogeneous turbulence and uniform particle distribution. 

As one can see, this equation is nonstationary (with a dependence on time). In his works, S.~Kaplan solved it analytically, which is a difficult and time-consuming task. To deal with it, he used Mellin transform (a version of the Laplace transform):
$$
H(\zeta,t)=\int\limits_{mc^2}^{\infty}\varepsilon^\zeta N(\varepsilon,t)d\varepsilon
$$
and showed that if the spectrum of accelerated particles at any time can be approximated by the expression $N(\varepsilon,t)\propto \varepsilon^{-(s(t)+1)}$, then it is obtained at $s(t)>1$: 
$$
s(t)=\frac{H(1,t)}{H(1,t)-mc^2 H(0,t)}, \quad N(t)=s(t)H(0,t).
$$
 
With this solution, S.~Kaplan analyzed the influence of the evolution of turbulence on the dynamics of the index $s(t)$ (e.g., [37]). In the work [30], he also estimated that the spectrum of cosmic rays at $t\rightarrow\infty$ should be $N(\varepsilon)\sim \varepsilon^{2.3}$, that is close to the observed values. 

It must be noted once again that S.~Kaplan related the acceleration of cosmic rays with the turbulent characteristics of the scattering centers where this acceleration occurs. Nowadays the main mechanism is believed to be a diffusion acceleration on the shock fronts. The kinetic equation for description of sych acceleration is obtained in different approaches. Both stationary and non-stationary solutions are of interest, in particular, nonlinear ones, which take into account the effect of accelerated particles on the flow structure (see reviews \cite{md2001,oprevcr}) or the growth of the magnetic field \cite{mfamplrev2012}. Numerical methods and powerful computing tools deepen our knowledge in cases where analytical solutions cannot be obtained, specifically in the joint solution of the equations of particle kinetics and turbulence, the formation of which they affect. Such effects have been actively studied in recent years by means of the so-called ``particle-in-cell'' simulations, which are based on the consideration of individual motions of a large number of particles in magnetized turbulence (see, for example, a review \cite{pnrev2020}). 

\section{Physics of the interstellar medium}

The main results of Samuil Kaplan’s work in the field of interstellar physics are presented in the book ``Physics of the interstellar medium'' \cite{ISM_1}, which he co-authored with S.~Pickelner. The book was published in 1979, after the author’s death. It provides basic information about gas, molecules, dust, cosmic rays, and the magnetic field in the interstellar medium. Some important aspects of the radiation transfer in the interstellar medium were also considered, and the data of the then observations in almost all ranges of the electromagnetic spectrum were reviewed. It should be noted that $\S$14, devoted to determining the ionization structure of interstellar gas, covers most of the basic principles of modern modeling of physical conditions in the interstellar medium. However, this chapter was not written by the authors of the book, but by N.~Bochkarev. In the chapters that follow, the authors describe the chemical processes in the interstellar medium and its dynamics, as well as the star formation processes in different galaxies. A sufficient part of the material published in the book is a generalization and development of S.~Kaplan's works of the Lviv period, which previously appeared on the pages of The Soviet Astronomical Journal [7,21-24,29,32,34,35,59,61,62,73], Reports of the USSR Academy of Sciences [10,72], Circulars of the Astronomical Observatory of the Lviv University [11,12,17,18,28,36,38,56,57,68,71], and four monographs [41,42,46,51].

One of the first Samuil Kaplan’s works on the study of the interstellar medium [12], published in 1952, is devoted to the development of a method for determining physical characteristics of the space dust. The interstellar absorption of the stellar radiation can be used to determine the dependence of the absorption coefficient on the wavelength, and the reflection of light by the dust nebulae can be used to estimate the dust scattering albedo as well as the dust scattering indicatrix. Based on these data, the author suggested calculating the refractive index of dust material, as well as its distribution by radii. To solve this problem, it is necessary to know the theoretical values of the absorption coefficient and the scattering indicatrix at the given values of the refractive indices. In modern calculations of the photoionization models of the gas-dust nebula luminescence (see \cite{ISM_Dust}) the absorption coefficient and scattering indicatrix are calculated using the Mie theory. In the aforementioned work, Samuil Kaplan offered to utilize the approximate method for estimating these characteristics of dust, based on simple analytical expressions of Shyfrin \cite{Shyfrin} and Van de Hulst \cite {vandehulst}. It was concluded that the proposed approximate method had an error of about 30\%. At that time, such accuracy of the theoretical estimates was considered acceptable given the accuracy of observational data and the simplest models of dust (sphericity and homogeneity of dust, in particular). It should be added that in 1953, S.~Kaplan also published papers on the condensation of interstellar gas on the dust particles [17], the interaction of interstellar hydrogen with the dust particles and their effect on the density of L$_\alpha$-radiation in the interstellar space [18].

In 1952, ``The Astronomical Journal'' published S.~Kaplan’s work [7] in which the theory of light scattering in the medium with absorption (developed by V.~Ambartsumian and V.~Sobolev) was used to study the reflection of light by the dust nebulae with infinitely large optical thickness. The nebulae were represented in the form of plane-parallel layers. The author also identified some characteristics of the space dust. S.~Kaplan’s work [57], published in 1953 in collaboration with I.~Klymyshyn, can be viewed as a continuation of this area of research. It considered the problem of light scattering (both central and diffuse) in a spherically symmetric nebula for a wide range of optical thicknesses. The author wrote an integral equation to determine the intensity of reflection of the unit area of the inner nebula surface of both radiation of the central star and radiation of the rest of the inner surface (integrated in all directions). In this case, to determine the reflection coefficient, the author used the expression obtained by V.~Ambartsumian \cite{ISM_Amba}. 
Segregation in the above-mentioned equation of the expression for the integral reflection coefficient allowed to calculate the value of the latter for different angles by means of numerical integration. Having these data, the author calculated the distribution of the intensity of diffuse radiation on the outer surface of the nebula. The expression used to calculate the transmittance coefficient was introduced by V.~Ambartsumian \cite{ISM_Amba}. Since computer technology was still in its infancy at that time, the author ``manually'' calculated the integrated reflection coefficient and the radiation intensity distribution from the outer surface of the nebula for different angles, as well as the optical thicknesses and albedo values, and presented the results in tables. These data were used to determine the density of diffuse L$_c$-radiation at the inner boundary of the nebula. The author had also graphically visualized the distributions of surface intensity over the visible surface of the nebula. These results can be used to compare with the observational data. This work is a vivid example of the ``pre-computer'' modeling of the nebulae luminescence. Thus, Samuil Kaplan can be seen as one of the pioneers of modern modeling of the luminosity of nebular environments.

The work [56], written in 1953 by S.~Kaplan in co-authorship with S.~Hopasiuk, is devoted to the study of the collisional excitation of interstellar hydrogen by electrons. The motivation for writing this work was G.~Shajn and V.~Gaze’s discovery \cite{Shain} of filament nebulae, near which there are no hot stars. The main mechanism of hydrogen radiation in the interstellar medium in the emission lines is its ionization with the subsequent recombination of electrons to the upper energy levels of hydrogen and the cascade transitions to lower levels. However, the lack of sources of ionizing radiation near the above-mentioned fibrous nebulae had forced astrophysicists to consider another way to populate hydrogen energy levels, i.e., collisions with electrons. J.~Oort \cite{Oort}  assumed that the collision of gaseous clouds in the interstellar medium could create conditions favorable for such a mechanism of excitation of hydrogen glow. S.~Kaplan and S.~Hopasiuk set out to test J.~Oort's hypothesis quantitatively. 

It is interesting to compare how the energy dependence of the effective cross section of excitation and ionization of hydrogen atoms by electron impact was determined in those days with how it is done now. Nowadays (see, for example, \cite{ISM_Rmatrix_HI}) to calculate the effective cross section of shock excitation and hydrogen deactivation, the $R$-matrix method \cite{ISM_Rmatrix}, is mainly used, which allows calculating the scattering parameters of some particles on others using the Schrödinger equation. S.~Kaplan and S.~Hopasiuk did not have such a toolkit at that time, so they used the empirical formula of V.~Fabrikant \cite{Fabrykant}, which divided the maximum value of this cross section for the excitation of the corresponding level and its energy dependence. The authors calculated the maximum value of this parameter relying on the Born method and following Bethe's calculations. According to the Born method, this value is proportional to the square of the matrix element of the corresponding transition. The authors calculated the number of excitations of hydrogen atoms in the first level on the basis of the energy dependence of the corresponding effective cross section for electrons, which, in turn, are distributed based on velocities according to the Maxwell distribution. Collisional deexcitation, however, was not taken into account. The rate of change in the number of protons per unit volume was determined by the difference between the amount of hydrogen ionization in the first level and recombination per unit time in a unit volume of the nebula. Recombination was taken into account starting from the second level, because recombination in the first level leads to the appearance of diffuse ionizing quanta, which, in the case of an optically thick interstellar medium, does not change the degree of ionization. Modern approaches to estimating this velocity (see, for example, sections 6.6–6.12 in the paper \cite{ISM_Hazy2}) consider in detail the scattering of photons of a given emission line, using the probabilities of photon escaping the nebula and its destruction, respectively. Such radiation contributes to the cooling of the interstellar medium, which must be paid attention to when modeling the radiation of the latter. Based on the aforementioned simplified approach, the authors [56] calculated the time of establishment of ionization-recombination equilibrium depending on the temperature and the degree of hydrogen ionization. 

The authors [56] also consider the Balmer decrement at different temperatures for two limit cases: 1) excitation by the electron impact, and 2) for the recombination spectrum. It should be noted that the system of stationary equations used by the authors does not include contributions from the collisional deactivations. The ratios of the intensities of the H$\alpha$/H$\beta$, lines give the minimum values for the recombination spectrum and the maximum values for the collisionally excited spectrum. It is worth mentioning that modern modeling of the luminescence of the environment surrounding the stars formation region (see, for example, \cite{ISM_MPhM}) allows us to take heed of both recombination and collisional contributions. The values of H$\alpha$/H$\beta$ calculated during such modeling are in the range of data obtained by modern observations \cite{ISM_Kehrig}. The same relations calculated by S.~Kaplan and S.~Hopasiuk in the aforementioned work also fit in the range of the observed values. Having analyzed the obtained results, the authors concluded that the radiation of interstellar hydrogen due to the excitation of its levels by electron impacts can occur only under conditions of constant heating of the medium. At the same time, turbulence was pointed out as a possible source of such heating. The authors highlighted the low efficiency of gas luminescence based on the passage of shock waves which result from the collision of clouds of the interstellar medium. It must be added that today the consideration of the collisional excitation of hydrogen luminescence in the emission lines is necessary not only during the precise photoionization modeling of the interstellar luminescence, but also during the diagnostic studies that require high accuracy. For example, G.~Stasinska and Yu.~Izotov in a study \cite{ISM_BCDG} of the effect of hydrogen collisional excitation on the determination of helium abundance in the low-metal HII regions, showed that consideration of this effect can change the ratio of H$\alpha$/H$\beta$ intensities to 8\% or more, which, in turn, affects the determination of the primordial helium abundance (which increases up to 5\%) synthesized in the era of cosmological nucleosynthesis. Thus, S.~Kaplan was one of the pioneers who seriously considered the effect of collisional excitation of hydrogen luminescence in the interstellar medium, which, as we can see, plays an important role in solving current problems of modern astrophysics.

The aforementioned turbulence of the interstellar cloud motion had been studied in detail by S.~Kaplan in the papers published in 1952 [12,71]. In [12] it was noted that the most reliable test of the hypothesis about the turbulent nature of the interstellar medium can be obtained from the analysis of the radial velocities of the gas clouds. In the same paper, a method for calculating the components of velocity correlation tensors was proposed, in which the relative radial velocity and the mutual distance of two clouds were directly compared. An attempt was also made to determine some components of the interstellar velocity correlation tensors with their mutual distances using two-component absorption lines H and K. In [71] the set of observed interstellar clouds was considered as a small heliocentric subsystem with a size of 200–500 pc. The authors looked for correlations between the average values of the clouds’ radial velocities (of different degrees) relative to the Sun and their distances to the Sun. Based on their results, the authors concluded that the hypothesis of the turbulent nature of interstellar gas motion was correct. A more detailed review of the studies of interstellar turbulence in S.~Kaplan's works can be found in Section 5 of this paper.

S.~Kaplan presented an intermediate summary of the approaches that use absorption lines to study the interstellar medium in 1957 [36]; in this publication, he demonstrated how based on the profiles of these lines one can 1) obtain information about the presence of individual clouds in the line of sight (by splitting these lines into components), 2) calculate the radial velocity of these clouds, or their average velocity (using the Doppler line width if no splitting is observed), 3) the dispersion of the velocities of atoms, ions and molecules utilizing the Doppler line width, 4) the number atoms, ions and molecules in a column of a single section along the line of sight of the observer to the star against which the line is observed (at full absorption). In fact, these approaches underlie the modern methods of analysis of the absorption lines formed in the interstellar medium.

We have reviewed only some of Samuil Kaplan's works on the physics of the interstellar medium written mainly during his work at the Lviv University. However, even such a brief overview demonstrates his enormous contribution to the given field of astrophysics. Samuil Kaplan is a founder of interstellar research at the Lviv University, where his scientific ``great-grandchildren'' and ``great-great-grandchildren'' continue to actively contribute to this field of study today.

\section{Physics and evolution of stars}
As we have noted above, Samuil Kaplan was known for the considerable comprehensiveness of his scientific interests. During the Lviv period, he wrote several scientific papers on the physics and evolution of stars, as well as the famous book ``Physics of stars'' [52], the first edition of which was published in 1961.

His first work on the physics of stars was [8], in which the author considered the calculation of the total energy emitted by the stars of a certain class in the interstellar medium. S.~Kaplan made an important refinement of similar calculations that had already existed at that time for the visible and ultraviolet parts of the spectrum. Despite the fact that the results obtained by S.~Kaplan differed from those observed at that time, the work is interesting because of its approach, which allows to calculate the flow based on the simplest principles. The discrepancies can be easily explained by the unavailability of the complete data on the spatial distribution of stars in the Galaxy and the distribution by classes of the stellar population at that time. This work was a valuable contribution to S.~Kaplan's favorite subject of interstellar physics since high-energy starlight is one of the decisive factors in it. Interestingly, at the end of the paper, the author thanked Joseph Shklovsky, one of the most prominent astronomers in the USSR, for discussing the issues presented in the paper. S.~Kaplan’s next work was devoted to the estimation of the level of star formation in a certain class of stars in the Galaxy [15]. At the time the paper was written, the scientific ideas concerning distribution of the number of stars by spectral classes in our Galaxy were far from being accurate, which makes this work significant only from a historical point of view. After all, it was only in the 1990s that the development of highly sensitive infrared astronomy allowed scientists to learn about the huge number of dim red and brown dwarfs, especially in the vicinity of the Sun. However, the calculation of the probability of formation of the stars of a certain class based on the observational function of luminosity and the law of evolution of stars of this class is relevant today, now such calculations are usually performed by a computer simulations. Another important paper on the evolution of stellar motion in the Galaxy is [5]. In this work, Samuil Kaplan successfully managed to estimate what he called the time of rotational relaxation of the Galaxy, namely, the time at which the Galaxy will enable a test body to reach the angular velocity that corresponds to a circular orbit around the center of the Galaxy through the gravitational interaction, at the initial velocity that appears to be very different from the corresponding angular velocity. This study showed the remarkable skill of Kaplan-theorist to work through estimates, rather than the accurate calculations of the physical quantities, as most formulas in the paper contain the sign $\approx$, not $=$. In addition, obtained results, namely the cubic dependence of this time on the distance to the center of the Galaxy, help to explain the strong homogeneity of the small elliptical galaxies. 

Finally, we must mention S.~Kaplan's famous book ``Physics of stars'', which was also written in Lviv. In a review of its English translation, Henny Lamers wrote, ``The book is written ... with a minimum of formulas, but with a deep understanding of physics and how different effects determine the structure and life of the stars. ... The book is very easy to read. It is written at a level that is appropriate for students, but anyone with the right knowledge of physics is able to comprehend it.'' As it is noted in the introduction to the book, knowledge on the level of school is sufficient to get a basic idea of the structure and evolution of stars, however, to build a model of processes, say, in a stellar atmosphere, such knowledge does not appear to be enough. On the other hand, ``Physics of stars'' addresses all the major issues of this subject, i.e., the basic parameters of stars, thermonuclear reactions, energy transfer in stars, classes, and types of stars and their evolution, compact astrophysical objects, variable stars, protostars, and star formation. The book contains a large amount of factual material; basically, it is a concentrated summary of physical knowledge about the stars at the time of writing. Thus, it served astrophysics students as a guide, and in some respects, it remains a valid source of information even today. The claim that the presentation of the material is really good might be confirmed by the fact that this book was republished. In 1977, a year before the tragic death of the author, the third supplemented and expanded edition of ``Physics of stars'' was published. The aforementioned English translation of ``Physics of stars'' appeared in 1982, after the scientist's death, published by John Wiley and Sons \cite{Kaplan1982}. 

Thus, during the Lviv period of his scientific work, Samuil Kaplan made a significant contribution to the study of physics of stars and their radiation in the interstellar space, as well as did outstanding pedagogical work to spark the interest of the general public in this subject.

\section{Cosmology and gravitation}

Samuil Kaplan's first work on cosmology entitled ``Cosmological turbulence'' was presented by the author himself at the sixth Conference on Cosmogony held in Moscow in 1957, June 5-7. This report was published in 1959 in the proceedings of the conference [49]. The publication took two printed pages and did not contain references. This work was based on Weizsäcker's interesting hypothesis about the origin of galaxies. According to this hypothesis, galaxies could have formed from the decay of the ``primary'' turbulence of the Metagalaxy into individual vortices. The impetus for this hypothesis was obviously the success in development of the theory of turbulence. In the given publication, S.~Kaplan did not express his attitude to the hypothesis but focused on its mathematical implementation. In particular, he wrote the equation of the evolution of turbulence in the case of the expanding Metagalaxy: 
\begin{eqnarray}
-\left(\frac{\partial}{\partial t}+2H\right) \int\limits_{0}^{k}F(k,t)dk =  
2\kappa\int\limits_{k}^{\infty} \sqrt{\frac{F(k, t)}{k^3}}dk\int\limits_{0}^{k}F(k,t)k^2dk + 2\zeta\int\limits_{0}^{k}  [F(k,t)k]^{3/2}dk,\label{eq8:1}
\end{eqnarray}
where $H$ is the Hubble constant (1/$H$ -- time of the Metagalaxy’s expansion), $F(k,t)dk$ -- vortex energy in the range of wavenumbers from $k$ to $k+dk$. 
The term with the multiplier $\kappa$ describes the energy transfer from the larger vortices to smaller ones, and the term with the multiplier $\zeta$ describes the dissipation of energy in metagalactic shock waves. With the introduction of new values
 $$y = \int\limits_{0}^{k}  \sqrt{\frac{F(k, t)}{k^3}}dk, \quad f =  \sqrt{\frac{F(k, t)}{k}}, 
 \quad \alpha = \frac{f}{y},$$
assuming that $\alpha$ does not depend on $k$ and $t$, the equation (\ref{eq8:1}) is rewritten as:
\begin{equation}
 \frac{1}{k^2}\left\{\alpha\frac{\partial}{\partial y}\frac{\partial y}{\partial t}
 - \frac{2}{y}\frac{\partial y}{\partial t} - (2 - \alpha)h \right\} = 
 (4-2\alpha)(\kappa+\zeta\alpha)-\kappa\alpha, \label{eq8:2}
\end{equation}
and its solution will be 
 $$f \sim k^{- \alpha}, \quad F \sim k^{1-2\alpha}.$$
Since the left-hand side of this equation depends on $k$, and the right-hand side does not, each of them equals zero, and, as a consequence, it is concluded that
$$F(k, t) = const\cdot e^{-2Ht}k^{1-2\alpha}, \quad \alpha = 1 - \frac34\frac{\kappa}{\zeta} + \sqrt{1+\frac12\frac{\kappa}{\zeta}+\frac{9}{16}\left(\frac{\kappa}{\zeta}\right)^2}.$$
This is the accurate solution of the equation (\ref{eq8:2}). Considering the case when dissipation in shock waves can be neglected (assuming $\zeta \ll \kappa$ and, consequently, $\alpha \rightarrow 4/3$), S.~Kaplan obtained the Kolmogorov spectrum with a time multiplier: $F(k, t) = const\cdot e^{-2Ht}k^{-5/3}$. In the case of dominance of the vortex energy dissipation in shock waves (assuming $\kappa \ll \zeta$ and, consequently, $\alpha \rightarrow 2$) S.~Kaplan got, as he noted, ``almost'' a stationary spectrum of vortices: $F(k, t) = const\cdot k^{-3}$.  

The problem of considering turbulence against the background of the homogeneous and isotropic universe is relevant today, given the indirect evidence of the existence of a primordial magnetic field that can generate turbulence in weakly ionized gas in the post-recombination universe.

Another Samuil Kaplan’s work devoted to cosmology is entitled ``Model of ``rotating'' space of constant curvature'' [54]; it has four printed pages and contains one reference to the recently published L.~Landau and E.~Lifshitz’s textbook ``Field theory''. At the end of the publication, it is noted that the calculations were performed by the fourth-year students of the Faculty of Physics of the Lviv State University, who majored in Theoretical Physics. S.~Kaplan and his students examined a model of the universe that rotates as a whole. The space-time interval of such a universe has the following general form:
\begin{eqnarray}
 ds^2 = a^2[d\eta^2-dx^2-\sin^2x(d\theta^2+\sin^2\theta d\phi^2)
 +2\sin^2{x} d\eta(\omega d\theta +  \sigma \sin^2\theta d\phi)],\nonumber
\end{eqnarray}
where $a$ is a radius of curvature of space, $\eta$ -- conformal time ($cdt=a(\eta)d\eta$), $x$ -- comoving radial distance, $\omega(\eta)$ and $\sigma(\eta)$ -- angular velocities of rotation of space at the angles $\theta$ and $\phi$, respectively. They calculated the nonzero components of the Einstein tensor, $R^k_j-\frac12\delta^k_j R$, which have the following form:
\begin{eqnarray}
 & R^0_0 - \frac12R = -\frac{3}{a^4}({\dot a}^2  a^2) + 2\frac{\dot a}{a}\ctg\theta\omega, \quad 
  R^1_1 - \frac12R = R^2_2 - \frac12R = -\frac{1}{a^2}\left[2\frac{\ddot a}{a}-\left(\frac{\dot a}{a}\right)^2+1\right] + \frac{\ctg\theta}{a^3}(a\dot\omega + 2\omega\dot a), \nonumber \\
 & R^3_3 - \frac12R = \frac{1}{a^2}\left[2\frac{\ddot a}{a}-\left(\frac{\dot a}{a}\right)^2+1\right], \quad
  R^1_0 = \frac{\omega}{a^2}\ctg\theta\ctg x, \nonumber \\
 & R^2_0 = \frac{2}{a^2}\left[2\frac{\ddot a}{a}-\left(\frac{\dot a}{a}\right)^2+1\right]\omega - \frac{\omega}{a^2\sin^2 x}, \quad
  R^3_0 = \frac{2}{a^2}\left[2\frac{\ddot a}{a}-\left(\frac{\dot a}{a}\right)^2+1\right]\sigma, \nonumber
\end{eqnarray}
where the dot indicates the derivatives with respect to conformal time $\eta$. The components of the stress-energy tensor have the following form:
\begin{eqnarray}
 &T^0_0 = -\rho c^2, \quad T^1_1=T^2_2=T^3_3=0, \quad T^1_0 = -\rho c^2 \frac{dx}{d\eta}, \quad
 &T^2_0 = -\rho c^2 \frac{d\theta}{d\eta}, \quad T^1_0 = -\rho c^2 \frac{d\phi}{d\eta},\nonumber
\end{eqnarray}
where $\frac{dx}{d\eta}$, $\frac{d\theta}{d\eta}$ and $\frac{d\phi}{d\eta}$ denote the systematic velocities that characterize the anisotropy of space, $\rho$ -- the average energy density of matter. Based on the Einstein's equations, solutions were found for the homogeneous rotating universe:
\begin{eqnarray}
  &a = const\cdot (1-\cos\eta), \quad
  \rho = \frac{const}{a^3} - \frac{\omega c^2}{4\pi G}\frac{\dot a}{a} \ctg{\theta},\quad
  \omega  = \frac{const}{a^2},\nonumber \\
  &\frac{dx}{d\eta} = - \frac{\omega c^2}{8\pi G\rho a^2}\ctg{x} \ctg{\theta},\quad
  \frac{d\theta}{d\eta} = - \frac{\omega c^2}{8\pi G\rho a^2}\cosec^2{x}.\label{eq8:3}\nonumber
\end{eqnarray}
The equations for $\sigma$ and $\frac{d\phi}{d\eta}$ are identically zero. The third equation concludes that the moment of rotation is preserved during the expansion. 

Due to the anisotropy of the model, the arbitrariness of the reference system is limited by the choice of azimuth $\phi$. A hypersurface with $\phi = const$ is a sphere with a linear element $d\psi^2 = dx^2+\sin^2{x}d\theta^2$, so the angular distance $\psi$ between a point on the sphere $\phi=0$ with the arbitrary coordinates $x$ and $\theta$ and the coordinates of our Galaxy on it $x_0$ and $\theta_0$ will be presented by the equation $\cos\psi = \cos{x}\cos{x_0} + \sin{x}\sin{x_0}\cos(\theta-\theta_0)$, while the linear distance will be equal to $a\psi$. The equation of propagation of light signals $d\eta^2=d\psi^2$, or $\psi = \pm \eta + const$, is the same as for the isotropic universe without rotation. The difference is the presence in the rotating universe of systematic velocities, which are different at different points in space. An observer in our Galaxy must detect a relative systematic velocity 
\begin{eqnarray}
 V = a\frac{d\psi}{dt} = \frac{d\psi}{d\eta}.\nonumber
\end{eqnarray}
In the general case, the dependence of the relative systematic velocity on $\psi$ and the direction to the observed galaxy is quite complex. Therefore, S.~Kaplan’s paper considered two extreme cases. In the first one, the observer sees a distant galaxy in the direction of the center/anticenter of the model. Then $\psi = x-x_0$, and consequently,
\begin{eqnarray}
 \frac{d\psi}{d\eta} = \frac{dx}{d\eta} - \frac{dx_0}{d\eta} =
 -\frac{\omega c^2}{8\pi G\rho a^2} \frac{\ctg \theta_0\sin\psi}{\sin x_0 \sin (x_0 + \psi)}.\nonumber
\end{eqnarray}
For short distances $\psi$ we observe that
\begin{eqnarray}
 V = \pm \frac{\omega c^3}{8\pi G\rho a^2} \frac{\ctg \theta_0}{\sin^2 x_0}\psi,\nonumber
\end{eqnarray}
where the sign ``$-$'' corresponds to the direction to the center, the sign ``$+$'', in contrast, to the anti-center. In this case, the systematic velocity like the Hubble velocity is proportional to the distance. In another extreme case, the galaxies in directions perpendicular to the center were considered, for which $\psi = \arccos (\cos x/ \cos x_0)$, and consequently,
\begin{eqnarray}
 \frac{d\psi}{d\eta} = 
 \frac{\omega c^2}{8\pi G\rho a^2} \frac{\cos\psi}{\sin\theta \sin\theta_0\sin x}.\nonumber
\end{eqnarray}
For short distances $\psi$ we see that
\begin{eqnarray}
 V = \frac{\omega c^3}{8\pi G\rho a^2} \frac{\cosec x_0}{\sin^2 \theta_0} = const.\nonumber
\end{eqnarray}
In this case, a stable addition to the rate of the Hubble expansion is obtained. According to the formulas presented in the work, one can calculate the value or the upper limit on the value of $\omega c^3/(8\pi G \rho a^2)$, and on the occasion of anisotropy, determine the direction to the center of the model. It should be noted that the question of deviation from the homogeneity and isotropy of the Universe on the scale of visible galaxies remains relevant today.

Another work on the theory of gravitation, this time a classic one, is entitled ``On the statistical theory of gravitational systems'' [66]; it is written in Ukrainian, takes three printed pages and contains five references. In this paper, Samuil Kaplan introduced the statistical function of the mass distribution $m_k$ based on the parameters of the solar orbit -- the major half-axis $a_k$, the eccentricity of the orbit $e_k$ and the inclination of the orbit $i_k$ utilizing the Boltzmann method. Using the integrals of the Lagrangian motion 
\begin{eqnarray}
 \sum\limits_k m_k\sqrt{a_k}e_k^2 = const, \quad 
  \sum\limits_k m_k\sqrt{a_k}\tg^2 i_k = const,\nonumber
\end{eqnarray}
which are valid for gravitationally interacting bodies that are in the outer field of a more massive body (the Sun), and the law of conservation of the total energy and mass of the system in the form of
\begin{eqnarray}
 \sum\limits_k \frac{m_k}{a_k} = const, \quad \sum\limits_k m_k = const,\nonumber
\end{eqnarray}
by finding the minimum of the function $H=\sum\limits_k m_k \ln m_k$, S.~Kaplan obtained the following distribution function:
\begin{eqnarray}
 m_k = C\exp\left\{-\sqrt{a_k}\left(\alpha e_k^2 + \beta \tg^2 i_k\right) - \frac{\gamma}{a_k}\right\},\nonumber
\end{eqnarray}
where $C$, $\alpha$, $\beta$ and $\gamma$ are the positive constants. From the above-mentioned distribution it could be concluded that the orbits with high eccentricity and/or inclination of the orbit are typical of the less massive bodies.

\section{Optical observations of the artificial Earth satellites}

In 1957, the Soviet Union started to prepare for the launch of the first artificial Earth satellite and began to organize a network of the optical observation stations. The latter was entrusted to the Astronomical Council of the USSR Academy of Sciences. Officially, there was no information about the preparation for the launch of the first satellite as all the work done was classified as ``top secret''. In the town of Firjuza, near the Iranian state border, during June-July 1957, the first all-union training courses for the heads of the satellite observation stations were organized at the Institute of Physics and Geophysics of the Academy of Sciences of Turkmenistan. Samuil Kaplan together with the famous astronomer Igor Astapovich were responsible for the course delivery. Future observers attended a course of lectures on the theory of motion of artificial Earth satellites and the methods of visual observations. At the same time, the observation stations of artificial satellites were set up at universities, pedagogical institutes, and astronomical observatories around USSR. At the time of the launch of the first satellite, 66 optical stations were ready for observations.

\begin{wrapfigure}{i}{0.5\textwidth}
\includegraphics[width=0.49\textwidth]{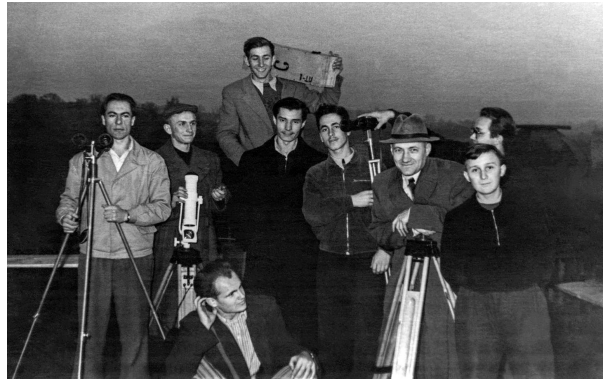}
\caption{S.~Kaplan with the students after observations of the first artificial Earth satellite (10 Oct 1957).}
\label{fig5}  
\end{wrapfigure}

In the Faculty of Physics of the Ivan Franko State University of Lviv, in the premises of the former meteorological station in the building in 8a Lomonosova Str. (now Kyryla and Methodia Str.) two rooms and a roof area were allocated for station. All the necessary equipment to record the passage of satellites during visual observations, i.e., astronomical optical tubes AT-1 (small wide-angle telescopes), radio receivers Riga-10, contact printing chronograph, binoculars, stopwatches, tape recorder, sound frequency oscillator, and Mikhailov’s star sky atlases were also provided. By the order of the rector of the University, the management of the station and the organization of observations of the first satellite were entrusted to S.~Kaplan. The university station became part of the network of optical observation stations of artificial satellites under number 1031; its work was coordinated by the Astronomical Council of the USSR Academy of Sciences.

Samuil Kaplan formed a group of future observers of the first artificial satellite and conducted classes with them on the methodology of visual observations. Among them there were Oleksandr Lohvynenko, Henadii Krainiuk and Yurii Frydel, who after graduating from the University formed the core of the future research department. The official TASS announcement of the launch of the first artificial Earth satellite PS-1 (Prosteyshiy Sputnik, that means the simplest satellite), was broadcasted on the radio on October 4, 1957, at 10 pm. However, cloudy weather in Lviv on the night of October 4-5, and in the next few days, did not allow the researchers to observe the movement of the satellite against the background of the star sky; the observers could only hear the signals on the radio. Only at night on October 10, 1957 (S.~Kaplan’s birthday), the first successful visual observations of the first artificial Earth satellite were made (Fig. \ref{fig5}). ``Analyzing the maximum volumes of the radio signals from the first satellite recorded on a tape recorder and comparing the intensity of these signals with the position of the satellite provided by the official newspaper ``Pravda'', S.~Kaplan noticed some discrepancy and realized that the ephemerides did not take into account the asphericity of the Earth,'' O.~Lohvynenko recollected \cite{Lohvynenko2011}. Having developed a method of graphical calculation of ephemeris, he solved this problem [40]. 

On November 3, 1957, the second largest artificial satellite Sputnik 2 was launched, which was easier to observe visually. Shortly afterwards, by the Resolution of the Presidium of the USSR Academy of Sciences No 854 of December 27, 1957, a photographic observation station of artificial Earth satellites was opened at the Astronomical Observatory of the Lviv University. The given station further functioned in a network of the similar stations under the name ``Lviv station of optical observations of artificial satellites 1031''. The Astronomical Council of the USSR Academy of Sciences provided scientific guidance, equipment, and consumables for all the stations in the Soviet Union. The Ephemeris Service provided the exact time and coordinates of the satellites which were sent with the help of special coded telegrams from the ``Cosmos'' Space Flight Control Center. Within the next two hours after the end of the observations, the results of the primary processing were sent in coded telegrams. With the increase in the number of launches of artificial satellites, there was a need to increase positional observations: visual observations were made by students and staff of the Faculty of Physics, while photographic ones were performed by the staff of the Astronomical Observatory. 

In November 1957, S.~Kaplan in co-authorship with I.~Astapovich published the first book in the history of world literature on the method of observation of the artificial Earth satellites, namely, ``Visual observations of artificial Earth satellites'' [40]. The book aimed at both professional observers and amateur astronomers in order to attract the maximum number of people to perform visual observations of the artificial satellites. The book presents basic information about the movement of the artificial Earth satellites, conditions of their visibility and methods of visual observation. The described methods allow the readers to approximately estimate the orbit of the satellite, calculate the moments of the satellite's passage over a specific geographical point, and obtain an estimate of its position in the sky at fixed points in time.
Samuil Kaplan stressed that visual-optical observations of the satellites are very important on the initial stages after launch, when the satellite's orbit is yet not known precisely. Knowing the exact orbit would solve such important problems as determining the structure of the Earth's crust and building a global geodetic network. To do this, it was necessary to determine the position of the satellite with an accuracy of several tens of meters in space. The book also presents the method of visual-optical observations of the meteors, which were studied by I.~Astapovich. This book has been translated into several languages, including Chinesse [95].

By the Order No 143 of the Ministry of Higher Education of the USSR issued on February 11, 1958, the study of satellites was included into the list of priority scientific research, and the Order No 729 (July 4, 1958) approved the ``Regulations on the optical observation stations of the artificial Earth satellites'', which governed the activities of the stations and laid all responsibility for their work on the heads of the institutions at which they were established. S.~Kaplan's efforts and his successful observations of the first and the subsequent satellites created favorable conditions for the development of hardware, the expansion of satellite observation methods, and the formation of a team of observers. The Astronomical Council of the USSR Academy of Sciences allocated three junior researcher positions; in addition, volunteers, who were mostly students, worked as observers, and were paid for successful visual observations.

Upon the initiative of the Astronomical Council of the USSR Academy of Sciences, special publications were issued, i.e., ``The Bulletin of the Optical Satellite Observation Stations'', ``Results of the Observations of the Soviet Artificial Earth Satellites'', etc. In 1958, the Astronomical Circular (USSR Academy of Sciences) published two scientific papers written by S.~Kaplan, where he presented the equation of motion of an artificial satellite and described his method of estimating the ephemeris of the satellites and their orbits [44,45]. At the same time, S.~Kaplan wrote another book ``How to see, hear, and photograph the artificial Earth satellites'' [43], which was recommended for professional observers as well as for radio amateurs and astronomy enthusiasts, with the goal to encourage them to observe independently and collaborate with the scientific institutions 
 
In the period from 1958 to 1961, S.~Kaplan together with I.~Klymyshyn were engaged in the scientific research devoted to the theory of satellite motion and the development of methods for predicting time and celestial coordinates of their passage for several days in advance. The second direction of their work was to modernize the equipment in order to increase the accuracy of observations and enable their automation, conducting a detailed study of instrumental errors. Both optical and radio engineering methods were used to determine the exact coordinates of the satellite to calculate its orbit. On the basis of these measurements, it was possible to obtain important data on the physical properties of the upper layers of the Earth's atmosphere. 

In general, S.~Kaplan supervised the station at 1957--1958, providing it with the necessary observation equipment, teaching the staff and students to observe fast-moving celestial objects. Samuil Kaplan made a significant contribution to the development of the theory of motion of artificial Earth satellites and the development of methods of optical observations and laid the foundation for further research in this area. Nowadays, the Astronomical Observatory of the Ivan Franko National University of Lviv has established a research complex with the equipment for the study of artificial celestial bodies in near space, which are the national Ukrainian heritage; it is also used to conduct laser ranging, photometric and positional observations of the artificial satellites. 

\section*{Acknowledgment}
The authors express deep gratitude to Andrii Rovenchak, Professor of the Department of Theoretical Physics named after Ivan Vakarchuk, for valuable help in forming the bibliography of S.~Kaplan’s works written during the Lviv period, insightful discussions and useful remarks. We would also like to thank the Armed Forces of Ukraine for their resilience and courage which saved us and  provided security to finalize this work during the unpvoked invasion of russian army into Ukraine.

\section*{Bibliography of S.~Kaplan's works written in the Lviv period\footnote{The papers 1-76 in this list were published in Russian which dominated in science in the USSR.}}
{\small
\begin{enumerate}
\item {\textit{Kaplan S.A.}} On circular orbits in Einstein's theory of gravitation (Letter to the editor). Zhurn. Exper. Theor. Phys. - 1949. - Vol. 19, No 10. - P. 951-952.
\item {\textit{Kaplan S.A.}} Energy sources and evolution of white dwarfs. Scientific notes (Lviv University). Phys.-math. Ser. Astronomy. - 1949. - Vol. 15, Issue 4. - P. 101-107.
\item {\textit{Kaplan S.A.}} Superdense stars. Scientific notes (Lviv University). Phys.-math. Ser. Astronomy. - 1949. - Vol. 15, Issue 4. - P. 109-116 .
\item {\textit{Kaplan S.A.}} Cooling of white dwarfs. Astron. Zhurn. - 1950. - Vol. 27, No 1. - P. 31-33.
\item {\textit{Kaplan S.A.}} The rotational relaxation time of the galaxy. Astron. Zhurn. - 1950. - Vol. 27, No 3. - P. 177-179.
\item {\textit{Kaplan S.A.}} About the article ``Time of rotational relaxation of the galaxy''. Astron. Zhurn. - 1950. -  No 3; (The author's letter to the editors about the erroneous conclusions with the editor's notes).  Astron. Zhurn. - 1950. - Vol. 27, No 5. - P. 320.
\item {\textit{Kaplan S.A.}} Reflection of light by dusty nebulae. Astron. Zhurn. - 1952. - Vol. 29, No 3. - P. 326-333.
\item {\textit{Kaplan S.A.}} The energy of the total radiation of stars. Astron. Zhurn. - 1952. - Vol. 29, No 6. - P. 649-653.
\item {\textit{Kaplan S.A.}} Brief report of the expedition of the Lvov Astronomical Observatory to observe the solar eclipse of February 25, 1952. Astron. circular (AS USSR). - 1952. - No 126. - P. 5-6.
\item {\textit{Kaplan S.A.}} On the conditions of an irrotational flow of gas in interstellar space. Rep. Acad. Sci. USSR. - 1952. - Vol. 87, No 6. - P. 909-912.
\item {\textit{Kaplan S.A.}} On the possibility of observational verification of the hypothesis of the turbulent nature of the motion of interstellar gas. Circular Astron. Observ. (Lviv University). - 1952. - No 23. - P. 1-5.
\item {\textit{Kaplan S.A.}} An elementary method for determining the physical characteristics of cosmic dust. Circular Astron. Observ. (Lviv University). - 1952. - No 23. - P. 5.
\item {\textit{Kaplan S.A.}} On the question of star formation. Astron. Zhurn. - 1953. - Vol. 30, No 4. - P. 391-393.
\item {\textit{Kaplan S.A.}} Quantitative characteristics of interstellar gas turbulence. Rep. Acad. Sci. USSR. - 1953. - Vol. 89, No 5. - P. 801-804.
\item {\textit{Kaplan S.A.}} On the question of star formation. Reports and announcements (Lviv University). - 1953. - Issue 4, ч.2. - P. 74-75.
\item {\textit{Kaplan S.A.}} Turbulence of interstellar gas. Reports at the 2nd meeting on cosmogony: in the book: Proceedings of the Second meeting on cosmogony. Moscow, 1953. - P. 221-224.
\item {\textit{Kaplan S.A.}} On the condensation of interstellar gas on cosmic dust particles. Scientific Notes (Lviv University). Ser. fiz.-mat. - 1953. - Vol. 22, No 5. - P. 111-114.
\item {\textit{Каплан С.А}}. Interstellar Hydrogen Interaction with Cosmic Dust and L$_{\alpha}$ Radiation Density in Interstellar Space. Scientific Notes (Lviv University). Ser. fiz.-mat. - 1953. - Vol. 22, No 5. - P. 115-120.
\item {\textit{Kaplan S.A.}} On the spectral function of isotropic turbulence in a compressible gas. Circular Astron. Observ. (Lviv University). - 1953. - No 25. - P. 1-4.
\item {\textit{Kaplan S.A.}} On the theory of acceleration of charged particles by turbulent magnetic fields. Circular Astron. Observ. (Lviv University). - 1953. - No 27. - P. 1-10.
\item {\textit{Kaplan S.A.}} Isothermal flow of gas in interstellar space. Density and speed jumps. Astron. Zhurn. - 1954. - Vol. 31, No 1. - P. 31-35.
\item {\textit{Kaplan S.A.}} Velocity distribution function of turbulent motion of interstellar gas. Astron. Zhurn. - 1954. - Vol. 31, No 2. - P. 137-170.
\item {\textit{Kaplan S.A.}} On the possibility of explaining the structure of filamentous nebulae. Astron. Zhurn. - 1954. - Vol. 31, No 4. - P. 358-359.
\item {\textit{Kaplan S.A.}} On conservation of velocity circulation in magnetogasdynamics. Astron. Zhurn. - 1954. - Vol. 31, No 4. — P. 360-361.
\item {\textit{Kaplan S.A.}} On the influence of stellar encounters on the correlation ``Eccentricity - Semi-major axis'': in the book: Questions of Cosmogony. 1954. - Vol. 2. - P. 269-272.
\item {\textit{Kaplan S.A.}} The system of spectral equations of magneto-gasdynamic isotropic turbulence. Rep. Acad. Sci. USSR. - 1954. - Vol. 94, No 1. - P. 33-37.
\item {\textit{Kaplan S.A.}} Spectral theory of gas-magnetic isotropic turbulence. Zhurn. Exper. Theor. Phys. - 1954. - Vol. 27, No 6. - P. 699-707.
\item {\textit{Kaplan S.A.}} On turbulent density fluctuations in interstellar space. Scientific notes (Lviv university). Astron. collection. - 1952. - Vol. 32, No 2. - P. 53-57.
\item {\textit{Kaplan S.A.}} Structural, Correlation, and Spectral Functions of Interstellar Gas Turbulence. Astron. Zhurn. - 1955. - Vol. 32, No 3. - P. 255-264.
\item {\textit{Kaplan S.A.}} On the theory of acceleration of charged particles by isotropic gas-magnetic turbulent fields. Zhurn. Exper. Theor. Phys. - 1955. - Vol. 29, No 4. - P. 106-116.
\item {\textit{Kaplan S.A.}} Shock waves in non-stationary stars: in the book: Proceedings of the Fourth Conference on Cosmogony. non-stationary stars. Moscow, 1955. - P. 178-185.
\item {\textit{Kaplan S.A.}} Shock waves in interstellar space. I. Astron. Zhurn. - 1956. - Vol. 33, No 5. - P. 646-653.
\item {\textit{Kaplan S.A.}} On the theory of the Fermi mechanism: in the book: Proceedings of the Fifth Conference on Cosmogony. Radio astronomy. Moscow, 1956. - P. 508-511.
\item {\textit{Kaplan S.A.}} Shock waves in interstellar space. ІI. Ionization discontinuities. Astron. Zhurn. - 1957. - Vol. 34, No 2. - P. 183-192.
\item {\textit{Kaplan S.A.}} Shock waves in interstellar space. ІІІ. Gasmagnetic discontinuities. Astron. Zhurn. -  1957. - Vol. 34, No 3. - P. 321-327.
\item {\textit{Kaplan S.A.}} Interstellar gas absorption lines. Circular Astron. Observ. (Lviv University). - 1957. - No 33. - P. 1-5.
\item {\textit{Kaplan S.A.}} Theory of statistical acceleration of charged particles by isotropic magnetic fields. Circular Astron. Observ. (Lviv University). - 1957. - No 33. - P. 6-25.
\item {\textit{Kaplan S.A.}} Nomograms for calculating gasmagnetic shock waves. Circular Astron. Observ. (Lviv University). - 1957. - No 33. - P. 26-28.
\item {\textit{Kaplan S.A.}} Methods of gas dynamics of the interstellar medium: Dr.Sci. dis. ... doc. phys.-math. Sciences: Moscow, 1957. 21 с. 
\item {\textit{Astapovich I.S., Kaplan S.A.}} Visual observations of artificial earth satellites. Moscow: Gostekhizdat, 1957. - 83 p. 
\item {\textit{Baum F.A., Kaplan S.A., Stanyukovich K.P.}} Introduction to space gasdynamics. Moscow: Fizmatgiz, 1958. - 424 p.
\item {\textit{Kaplan S.A.}} Interstellar gasdynamics. Moscow: Fizmatgiz, 1958. - 194 с.
\item {\textit{Kaplan S.A.}} How to see, hear and photograph artificial satellites of the Earth. Moscow: Fizmatgiz, 1958. - 80 p.
\item {\textit{Kaplan S.A.}} Equation of Motion of Artificial Earth Satellites and Control of Observations. Astron. circular (AS USSR). - 1958. - No 189. - P. 1-3.
\item {\textit{Kaplan S.A.}} Method for Approximate Calculation of Ephemerides of Artificial Earth Satellites and Determination of their Orbits. Astron. circular (AS USSR). - 1958. - No 192. - P. 5-8.
\item {\textit{Kaplan S.A.}} Magnetic gasdynamics and questions of cosmogony: in the book: Questions of cosmogony. Moscow, 1958. - P. 238-264.
\item {\textit{Kaplan S.A.}} On Cosmic Forceless Fields. Astron. Zhurn. - 1959. - Vol. 36, No 5. - P. 800-806.
\item {\textit{Kaplan S.A.}} On the Larmor theory of plasma. (Letter to the editor). Zhurn. Exper. Theor. Phys. - 1959. - Vol. 36, No 6. - P. 1927-1928.
\item {\textit{Kaplan S.A.}} Cosmological turbulence: in the book: Proceedings of the sixth conference on cosmogony. Extragalactic astronomy and cosmology. Moscow, 1959. - P. 260-262.
\item {\textit{Kaplan S.A.}} Influence of anisotropy of conductivity in a magnetic field on the structure of a shock wave in magnetic gas dynamics. (Letter to the editor). Zhurn. Exper. Theor. Phys. - 1960. - Vol. 38, No 1. - P. 252-253.
\item {\textit{Kaplan S.A.}} New data on outer space. Kyiv, 1960. - 37. p.
\item {\textit{Kaplan S.A.}} Physics of stars. Moscow: Fizmatgiz, 1961. - 151 p.
\item {\textit{Kaplan S.A.}} Simple waves and the formation of shock waves in stars. Circular Astron. Observ. (Lviv University). - 1962. - No 37-38. - P. 3-8.
\item {\textit{Kaplan S.A.}} Model of ``rotating'' space of constant curvature. Circular Astron. Observ. (Lviv University). - 1962. -  No 37-38. - P. 9-12.
\item {\textit{Kaplan S.A.}} About one magnetohydrodynamic problem. Bulletin of Lviv University. Phys. ser. - 1962. - Issue 1(8). - P. 73-74.
\item {\textit{Kaplan S.A., Gopasyuk S.I.}} Excitation of interstellar hydrogen glow by electron impacts. Circular Astron. Observ. (Lviv University). - 1953. - No 25. - P. 5-14.
\item {\textit{Kaplan S.A., Klimishin I.A.}} Scattering of light in spherical nebulae. Circular Astron. Observ. (Lviv University). - 1953. - No 27. - P. 11-16.
\item {\textit{Kaplan S.A., Klimishin I.A.}} On the limiting density of white dwarfs. Circular Astron. Observ. (Lviv University). - 1953. - No 27. - P. 17-22.
\item {\textit{Kaplan S.A., Klimishin I.A.}} On the correlation of the observed difference in the degree of interstellar polarization with the angular distance of the corresponding points on the celestial sphere. Astron. Zhurn. - 1959. - Vol. 36, No 2. - P. 370-371.
\item {\textit{Kaplan S.A., Klimishin I.A.}} Shock waves in the shells of stars. Astron. Zhurn. - 1959. - Vol. 36, No 3. - P. 410-421.
\item {\textit{Kaplan S.A., Klimishin I.A.}, Sivers V.N.} Theory of light scattering in a medium with a moving boundary. Astron. Zhurn. - 1960. - Vol. 37, No 1. - P. 9-15.
\item {\textit{Kaplan S.A., Klimishin I.A.}} Some Remarks on Light Emission by Shock Waves in Space Conditions. Astron. Zhurn. - 1960. - Vol. 37, No 2. - P. 281-283.
\item {\textit{Kaplan S.A., Klimovska A.I.}} On the equation of motion of an artificial Earth satellite in horizontal coordinates. Bulletin of stations for optical observations of artificial satellites of the Earth. - 1960. - No 1. - P. 10-12.
\item {\textit{Kaplan S.A., Kovalchuk V.G.}, Korolyshyn B.M.} Coefficients of electrical conductivity and diffusion in a relativistic one-component plasma. Bulletin of Lviv University. Phys. ser. - 1962. - Issue 1(8). - P. 79-82.
\item {\textit{Kaplan S.A.}, Kolodij B.I.} Functional alignment of magnetic hydrodynamics. Reports and announcements (Lviv University). - 1957. - Issue 7, ч.3. - P. 229-230.
\item {\textit{Kaplan S.A.}, Kravchuk A.N.} Before the statistical theory of gravitational systems. Reports and announcements (Lviv University). - 1957. - Issue 7, ч.3. - P. 230-232.
\item {\textit{Kaplan S.A.}, Kurt V.G.} Expansion of a sodium cloud in interplanetary space. Astron. Zhurn. - 1960. - Vol. 37, No 3. - P. 536-542.
\item {\textit{Kaplan S.A.}}, Kutyk I.N. About the development of magnetohydrodynamic and magneto sound waves. Bulletin of Lviv University. Phys. ser. - 1962. - Issue 1(8). - P. 75-78.
\item {\textit{Kaplan S.A., Logvinenko A.A., Podstrigach T.}} Before calculating the parameters of gas-magnetic shock waves. Ukr. fiz. zhurn. - 1959. - Vol. 4,  N4. - P. 438—447.
\item {\textit{Kaplan S.A., Logvinenko A.A., Porfir'ev V.V.}} Some questions of magnetic gas dynamics. Questions of magnetic gas dynamics and plasma dynamics: collection of reports. Riga owl. Riga, 1962. - P. 33-37.
\item {\textit{Kaplan S.A., Pronik V.I.}}  On the question of interstellar gas turbulence. Circular Astron. Observ. (Lviv University). - 1952. - No 24. - P. 1-5.
\item {\textit{Kaplan S.A., Pronik V.I.}} On the question of the turbulent nature of the motion of interstellar gas clouds. Rep. Acad. Sci. USSR. - 1953. - Vol. 89, No 4. - P. 643-66.
\item {\textit{Kaplan S.A.}, Sivers V.N.} General problem of light scattering in a one-dimensional medium with a moving boundary. (To the problems of astrophysics. Astron. Zhurn. - 1960. - Vol. 37, No 5. - P. 824-827.
\item {\textit{Kaplan S.A., Staniukovich K.P.}} Solving the equations of magneto-gasdynamics for one-dimensional motion. Rep. Acad. Sci. USSR. - 1954. - Vol. 95, No 4. - P. 769-771.
\item {\textit{Kaplan S.A., Staniukovich K.P.}} On the solution of inhomogeneous problems of one-dimensional motion in magnetic gas dynamics. Zhurn. Exper. Theor. Phys. - 1956. - Vol. 30, No 2. - P. 382-385.
\item {\textit{Kaplan S.A., Tsap T.T.}} Ionization functions of elements KI, CaI, CaII, NaI, CI in interstellar space. Astron. circular (AS USSR). - 1953. - No 137. - P. 6-7.
\item {\textit{Kaplan S.A.}} The theory of the acceleration of charge by isotropic gas magnetic turbulent fields. Soviet Phys. JETP. - 1956. - Vol. 2, No 2. - P. 203-210.
\item {\textit{Kaplan S.A., Staniukovich K. P.}} The Solution of Inhomogeneous, One-Dimensional Motion Problems in Magnetic Gasodynamics. Soviet Phys. JETP. - 1956. - Vol. 3, No 2. - P. 275-277.
\item {\textit{Kaplan S.A.}} Shock Waves in Interstellar Space. II. Ionization Discontinuities. Soviet Astron. - 1957. - Vol. 1. - P. 183-191.
\item {\textit{Kaplan S.A.}} Shock Waves in Interstellar Space. III. Hydromagnetic Discontinuities. Soviet Astron. - 1957. - Vol. 1. - P. 317-323.
\item {\textit{Kaplan S.A.}} Shock Waves in Magnetogasodynamic Turbulence. Rev. Mod. Phys. - 1958. - Vol. 30, No 3. - P. 1089.
\item {\textit{Kaplan S.A.}} On the formation of the interstellar gas clouds. Rev. Mod. Phys. - 1958. - Vol. 30, No 3. - P. 943.
\item {\textit{Kaplan S.A.}} Theory of Isotropic Magnetic Turbulence in Gases. Electromagnetic Phenomena in Cosmical Physics. IAU Symposium no. 6: proceedings / Ed. Bo Lehnert.: Cambridge University Press. - 1958. - P. 504-512.
\item {\textit{Kaplan S.A.}} Jak vid\v et, sly\v set a fotografovat um\v ele dru\v zic\'{e} Zem\v e. Praha, SNTL. 1959. - 89 s.
\item {\textit{Kaplan S.A.}} On Cosmic Force-Free Fields. Soviet Astron. - 1959. - Vol. 3.— P. 778-783.
\item {\textit{Kaplan S.A.}} Effect of conductivity anisotropy in a magnetic field on the structure of a shock wave in magnetogasdynamics. Soviet Phys. JETP. - 1960. - Vol. 11, No 1. — P. 183.
\item {\textit{Kaplan S.A.}} Shock-Waves in Stars. Mod\'{e}les d’Etoiles et \'{E}volution Stellaire: communications pr\'{e}sent\'{e}es au neuvi\`{e}me Colloque International d'astrophysique, tenu \`{a} Li\`{e}ge, les 6, 7 et 8 juillet 1959.— Li\`{e}ge: Soci\'{e}t\'{e} Royale de Li\`{e}ge, 1960. — P. 296-297.
\item {\textit{Kaplan S.A.}} On the theory of propagation and decay of hydromagnetic waves in the anisotropic medium. Rev. Mod. Phys. 1960. - Vol. 32, No 4. - P. 881.
\item {\textit{Kaplan S.A., Klimishin I.A.}} On the Correlation Between the Observed Difference in the Degree of Interstellar Polarization and the Angular Distance of the Corresponding Points on the Celestial Sphere. Soviet Astron. - 1959. - Vol. 3. - P. 362.
\item {\textit{Kaplan S.A., Klimishin I.A.}} Shock waves in stellar envelopes. Soviet Astron. - 1959. - Vol. 3. - P. 404-414.
\item {\textit{Kaplan S.A., Klimishin I.A.}} Some Notes on the Emission of Light by Shock Waves under Cosmic Conditions. Soviet Astron. - 1960. - Vol. 4. - P. 264-266.
\item {\textit{Kaplan S.A., Klimishin I.A.}, Sivers V.N.} The Scattering of Light in a Medium with a Moving Boundary. Soviet Astron. - 1960. - Vol. 4. - P. 7-12.
\item {\textit{Kaplan S.A., Kurt V.G.}} Expansion of a Sodium Cloud in Interplanetary Space. Soviet Astron. - 1960. - Vol. 4. - P. 508-514.
\item {\textit{Kaplan S.A., Sivers V.N.}} The General Problem of the Scattering of Light in a One-Dimensional Medium with a Moving Boundary. Soviet Astron. - 1960. - Vol. 4. - P. 776-779.
\item {\textit{Astapovich I.S. Kaplan S.А.}} Visual observation of artificial Earth satellites. -- Beijing: Science Press, 1959. -- 73 p. (in Chinesse).

\end{enumerate}
}

\end{document}